\documentclass[aps,twocolumn,pra,superscriptaddress,amsmath,showpacs,tightenlines]{revtex4-2}
\usepackage{epsfig,graphicx,times}
\usepackage{amstext}
\usepackage{amsmath}            %serve per le subequazioni
\usepackage{amssymb}            %serve per il simbolo "marchio registrato", \circledR
\usepackage{graphicx}           %serve per le figure eps, ps etcyy
\usepackage{latexsym}
\usepackage{color}
\usepackage{bm}

\begin{document}

\title{Control the qubit-qubit coupling with  double superconducting resonators}

\author{Hui Wang}
\email{wanghuiphy@126.com}
\affiliation{Department of Physics, Graduate School of Science,
Tokyo University of Science, Shinjuku-ku, Tokyo, Japan}
\affiliation{RIKEN Center for Quantum Computing (RQC),Wako-shi, Saitama, Japan}

\author{Rui Wang}
\affiliation{Fujitsu Limited, Nakahara-ku, Kawasaki, Kanagawa 211–8588, Japan}

\author{Daichi Sugiyama}
\affiliation{Department of Physics, Graduate School of Science,
Tokyo University of Science, Shinjuku-ku, Tokyo, Japan}

\author{Chih-Yao Shih}
\affiliation{Department of Electrophysics, National Yang Ming Chiao Tung University, Hsinchu 300093, Taiwan}
\affiliation{RIKEN Center for Quantum Computing (RQC),Wako-shi, Saitama, Japan}

\author{Ching-Yeh Chen}
 \affiliation{Department of Physics, National Tsing Hua University, Hsinchu 30013, Taiwan}
\affiliation{RIKEN Center for Quantum Computing (RQC),Wako-shi, Saitama, Japan}

\author{Hiroto Mukai}
\affiliation{RIKEN Center for Quantum Computing (RQC),Wako-shi, Saitama, Japan}

\author{Hang Xue}
\affiliation{RIKEN Center for Quantum Computing (RQC),Wako-shi, Saitama, Japan}

\author{J.S. Tsai}
\affiliation{Department of Physics, Graduate School of Science,
Tokyo University of Science, Shinjuku-ku, Tokyo, Japan}
\affiliation{RIKEN Center for Quantum Computing (RQC),Wako-shi, Saitama, Japan}
\affiliation{Department of Electrophysics, National Yang Ming Chiao Tung University, Hsinchu 300093, Taiwan}

\date{\today}

\begin{abstract}
We experimentally studied the switching off processes in the double-resonator coupler superconducting quantum circuit.
In   both  frequency  and time-domain, we observed the variation of qubit-qubit effective coupling by tuning the
 frequency differences between qubits and the double-resonator coupler.
 According to the measurement results, by just shifting about 50 MHz of  qubits' frequencies,
 we can tune the effective qubit-qubit coupling strength from   switching off point  to two qubit gate point (effective coupling larger than 5 MHz)   in double-resonator  superconducting quantum circuit.
 The double-resonator (coupler) superconducting quantum circuit
  has the advantage of    simple fabrications, introducing less flux noises,
    reducing occupancy of  dilution refrigerator cables,
 which might supply a promising platform for future large-scale superconducting quantum processors.
  \end{abstract}

\maketitle
 \pagenumbering{arabic}

\section{Introduction}

The tunable coupler has the advantage of switchable interaction, fast two-qubit gates,reduced residual ZZ and idling errors, and cleaner scaling to many qubits, which   becomes indispensable devices in the multi-qubit superconducting quantum chip\cite{Chen,Yan,Sun,Tan1,Martinis,Moskalenko,Wu,Sung}.  The transmon coupler is the mainstream in the current large-size superconducting quantum chip and realizing the quantum supremacy and initial quantum errors \cite{Martinis,Moskalenko,Wu,Sung,google1,google2,Tan}.

 As shown by previous  theoretically and experimentally research, the superconducting resonator  can also function as a  tunable coupler for the superconducting qubits\cite{wang,Wang1,Wu2,Ming,IBM,McKay,Stehlik}.  According to recent theoretical work \cite{Wang1}, the double-resonator coupler can switch off the qubit-qubit effective coupling and their static ZZ coupling, and the double-resonator coupler have special  advantage of simple fabrications, small frequency tuning ranges,  less flux noise,   and without occupying a  dilution refrigerator cables.

In this article, we experimentally studied the switching off processes in the double-resonator coupler superconducting quantum circuit. In   both  frequency  and time-domain, we measured the variation of qubit-qubit coupling strength by tuning frequency differences of qubits relative to the resonators coupler  through the Z-control signals.
By changing the energy level crossing points relative to the fixe frequency resonator  with the Z-control signals (DC-bias current),  we observed the variation of qubit-qubit anti-crossing gap  with two-tone spectroscopy on superconducting quantum circuits.
Even with the relative noise measurement data in time-domain measurement(without Josephson parameter amplifier and  high base-temperature of dilution refrigerator above  25mK), we still can see the variation of envelope of vacuum Rabi oscillation which reflects the change of effective qubit-qubit coupling. According to the measurement results,  the effective qubit-qubit coupling can be tuned from switching off point to two qubit gate point (effective coupling larger than 5 MHz) by just shifting about 50 MHz of  qubits' frequencies in double-resonator  superconducting quantum circuit.

The paper is organized as follows:  In Sec. II,  we theoretically analyze the effective  qubit-qubit coupling. In Sec. III, the two tone measurement in frequency domian . In Sec. IV,  the time-domain measurement of effective qubit-qubit coupling. Finally, we summarize the results in Sec. V.

\section{Theoretical model}

\begin{figure}
\includegraphics[bb=130 10 900 700, width=7 cm, clip]{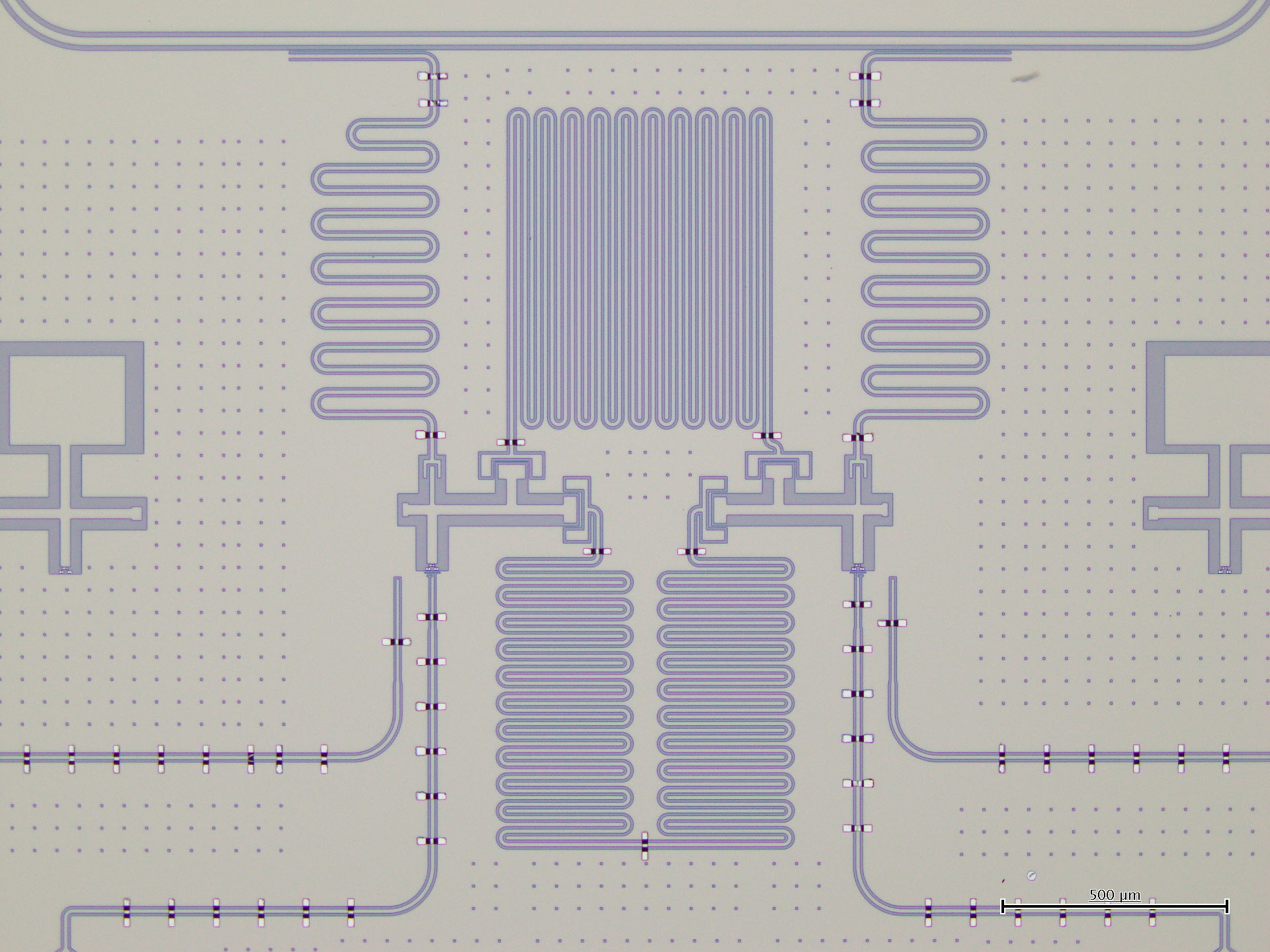}
\caption{(Color online) Optical image for Superconducting quantum circuits used in
our experiment. The chip consists of two frequency tunable qubits
 coupling to two common  superconducting resonators.
The independent XY-control (or Z-control)  line is used
for microwave-driven (or frequency tuning) of corresponding qubit .
The two-body interactions are mainly   capacitive types.
}
\label{fig1}
\end{figure}

Figure~\ref{fig1} show the sample of double-resonator superconducting quantum circuit,
 which  consist of two tunable Xmon qubits coupling to two common fixed frequency resonators.
 The resonant frequencies $\omega_{a,b}$ of resonator \textbf{a,b} are fixed,
and the  transition frequencies $\omega_{1,2}$ of qubit-\textbf{1,2}  can be tuned by the
independent Z-control signals.  In future multi-qubit chip, the size of resonator coupler   can be reduced by
  using narrower coplanar waveguide resonators to reduce the  crosstalks.

According to the previous theoretical work,
 the second-quantization  Hamiltonian of the double resonator coupler superconducting circuit in Fig.1(a) can be written as \cite{Wang1}
 \begin{eqnarray}\label{eq:3}
 H&=&\frac{\hbar}{2} \sum_{\lambda=a,b}{ \omega_{\lambda} c^{\dagger}_{\lambda} c_{\lambda}}+\frac{\hbar}{2}\sum_{\beta=1,2}{\left(\omega_{\beta} a^{\dagger}_{\beta} a_{\beta}+ \alpha_{\beta} a^{\dagger}_{\beta} a^{\dagger}_{\beta} a_{\beta} a_{\beta}\right)}\nonumber\\
 &+ &\hbar \sum_{\lambda=a,b \atop \beta=1,2}{ g_{\lambda\beta}(c^{\dagger}_{\lambda}a_{\beta}+c_{\lambda}a^{\dagger}_{\beta}-c^{\dagger}_{\lambda}a^{\dagger}_{\beta}-c_{\lambda}a_{\beta})}\nonumber\\
 & +&\hbar g_{ab}(c^{\dagger}_{a}c_{b}+c_{a}c^{\dagger}_{b}-c^{\dagger}_{a}c^{\dagger}_{b}-c_{a}c_{b})\nonumber\\
 &+ &\hbar g_{12}(a^{\dagger}_{1}a_{2}+a_{1}a^{\dagger}_{2}-a^{\dagger}_{1}a^{\dagger}_{2}-a_{1}a_{2}).
 \end{eqnarray}
 The  transition frequencies of resonators and qubits are respectively defined as $\omega_{\lambda}$ and $\omega_{\beta}$, while  $\alpha_{\beta}$ labels the anharmonicity of qubit $\beta$, with $\lambda=a,b$ and $\beta=1,2$. $g_{\lambda\beta}$ is the  coupling strength between resonator-${\lambda}$ and qubit-${\beta}$,  while $g_{12}$ (or $g_{ab}$) describe the directly qubit-qubit (or resonator-resonator)  capacitive coupling.

The qubit-resonator coupling strength $g_{\lambda\beta}$ can induce  indirect  interactions between two qubits,
and the effective qubit-qubit coupling strength can be obtained as
\begin{eqnarray}\label{eq:5}
g_{eff}=\frac{1}{2}\sum_{\lambda=a,b \atop \beta=1,2}\left(\frac{g_{\lambda 1}g_{\lambda 2}}{\Delta_{\lambda \beta}}-\frac{g_{\lambda 1}g_{\lambda 2}}{\Sigma_{\lambda \beta}}\right)+g_{12}.
\end{eqnarray}
Where we have defined $\Delta_{\lambda\beta}=\omega_{\beta}-\omega_{\lambda}$ and $\Sigma_{\lambda\beta}=\omega_{\beta}+\omega_{\lambda}$.
According to Eq.(2),  the effective qubit-qubit coupling $g_{eff}$ can be tuned by changing qubits' frequencies $\omega_{\beta}$ through the flux signals applying from corresponding Z-control lines.

If the transition frequencies of  two qubits are tuned between that of two   resonator coupler ($\omega_b>\omega_{1,2}>\omega_a$),
the high and low frequency resonators  will respectively  make a negative and positive contributions to the qubit-qubit interaction.
When the net contribution of two resonators is same value but opposite sign with the direct qubit-qubit coupling $g_{12}$, the qubit-qubit coupling can be switched off.
Even for extreme  small (or even zero) value of $g_{12}$, the switching off ($g_{eff}=0$) can be realized without requiring   large-scale
 of  frequency shift in the  double-resonator coupler superconducting quantum circuit.

\section{frequency domain measurement}

The two tone spectroscopy for double-resonator coupler superconducting quantum circuit in Fig.1 are measured with network analyzer (keysight N5231B),
  microwave source (Agilent E8247c), and two DC-current sources (GS200).
 With the   DC-bias current, according to the  energy spectrum  in the Fig.2(a),
 the DC-bias frequency of qubit-\textbf{1} (qubit-\textbf{2}) is about 4.641 GHZ  (or 4.691 GHz).
 Under  strong microwave driving field, the energy spectrum of fixed frequency  resonator couplers (4.47 GHz and 4.80 GHz) and
 second-excited states of qubits can also be observed  as  shown in Fig.2.
  Since maximal frequency
   of qubit-\textbf{1} is  between resonant frequencies of  two   resonator-coupler ($\omega_b> \omega^{(max)}_1> \omega_a$),
   thus we only see the anti-crossing gap of qubit-\textbf{1} with  resonator-\textbf{a} in Fig.2(a).
    Since the maximal frequency of qubit-\textbf{2} is larger than the  resonator frequency of two  resonator couplers ($\omega^{(max)}_2>\omega_b>\omega_a$),
    so we can observe the anti-crossing gap of qubit-\textbf{2} with  resonator-\textbf{a}
    and resonator-b under the DC-bias current as shown in  Fig.2(b).
  The effective coupling of two qubits with the low frequency resonator
  are about 28 MHz, while 30 MHz with the high frequency resonator according to
  the measurement energy spectrum or  design simulation (qubit-\textbf{1} and resonator-\textbf{a}).

\begin{figure}
\includegraphics[bb=10 0 630 500, width=6 cm, clip]{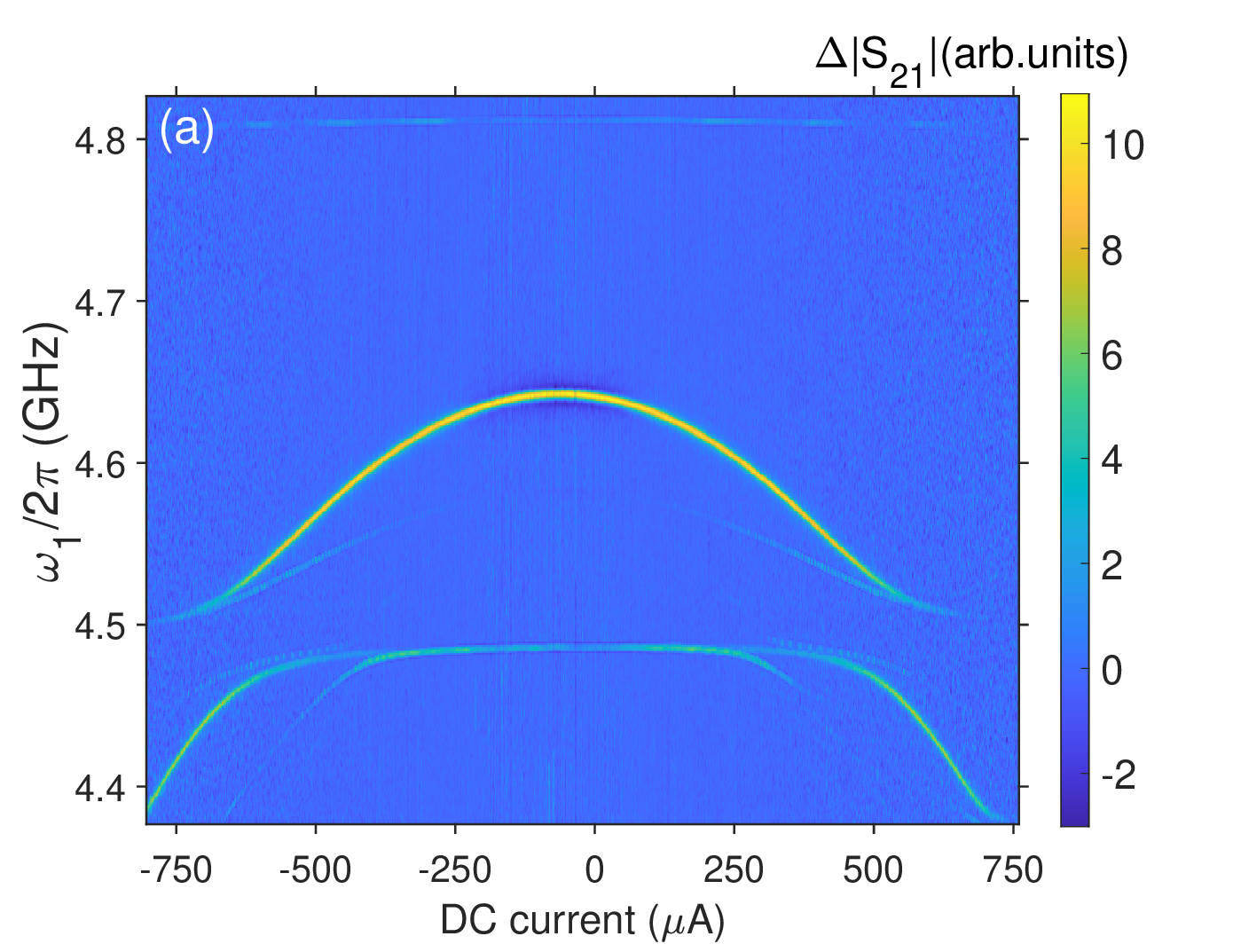}\\
\includegraphics[bb=0 0 640 530, width=6.2 cm, clip]{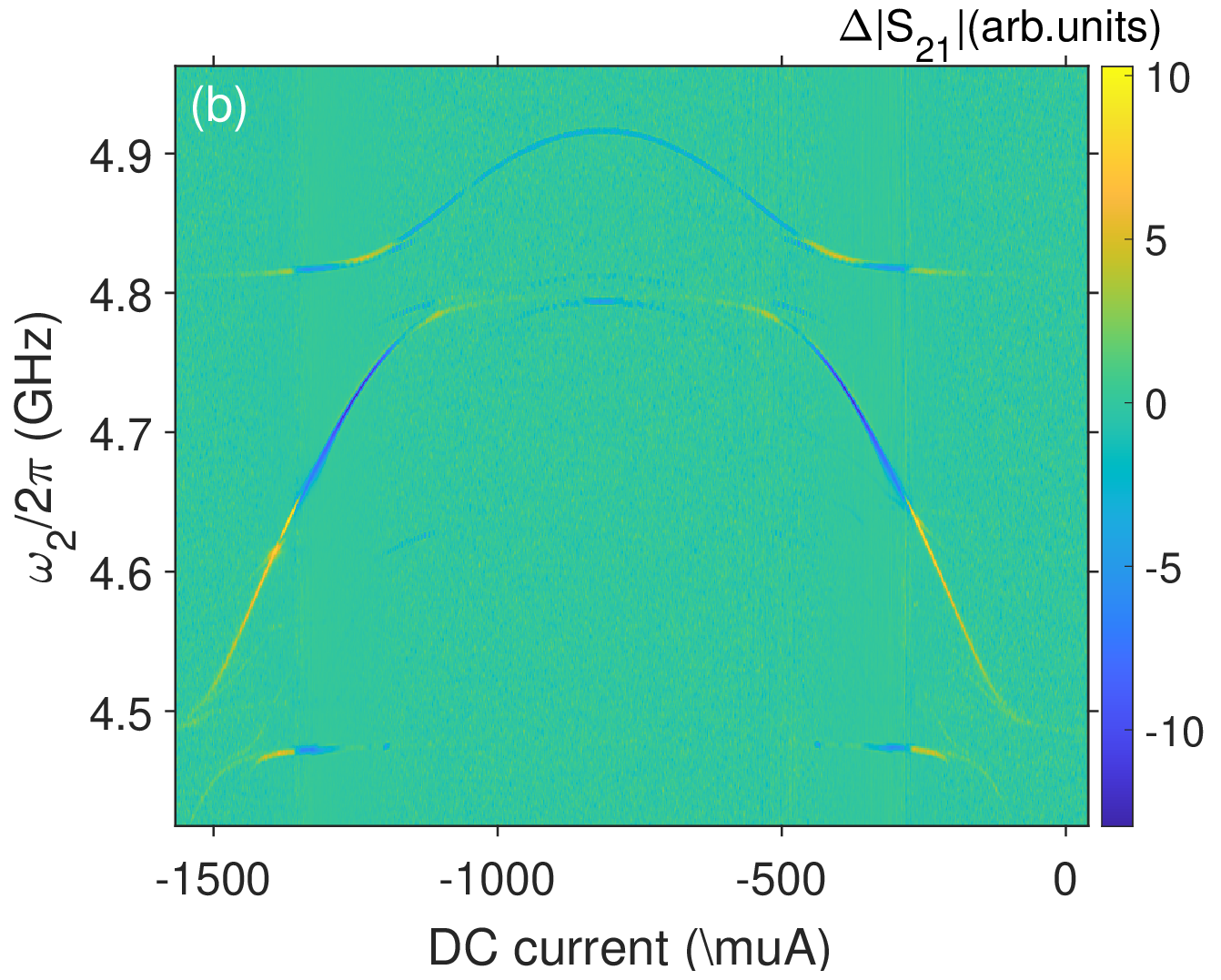}
\caption{(Color online) Energy spectrum of qubits under the DC-bias  current.
The maximal frequency of qubit-1 and qubit-2  are about 4.641 GHz and 4.91 GHz, respectively.
 Under the strong pumping field, the energy levels of  resonator couplers and
 second-excited states of qubits can also be seen in the energy spectrum.
The resonant frequency of low-frequency  resonator-a and high-frequency is about 4.47 GHz
 and 4.80 GHz, respectively. According to corresponding anti-crossing gap,  the
   coupling strength of high (or low)-frequency resonator with qubits  are about 30 MHz (or 27 MHz).
}
\label{fig2}
\end{figure}

The transition frequency  of qubit--\textbf{2} is fixed at the certain  value  by DC-bias current,
then the effective qubit-qubit coupling depends on the frequency of qubit-\textbf{1} (or $\Delta_{\lambda\beta}$) according to Eq.(2).
 If we apply an independent bias current  to sweep  the frequency of qubit-\textbf{1} with DC-biased frequency around transition frequency of qubit-\textbf{2},
 then  the variation of qubit-qubit anti-crossing gap are shown in  Fig.3.
 By fixing the frequency of qubit-\textbf{2}  at different frequency with DC bias current, and the frequency of qubit-1 sweep around the frequency of qubit-\textbf{2}.
The anti-gap reduce from about 10 MHz (4.58 GHz) in Fig.3(a) and reduce to below 2 MHz in Figs.3(d) and  3(e) (close to 4.62 GHz).
 If  qubit-\textbf{2} is tuned to about 4.37 GHz, the anti-crossing gap is almost invisible in Fig.3(e) where the qubit-qubit coupling is turned off.
By choosing direct qubit-qubit coupling as 0.88 MHz, with the qubit-resonator coupling strength obtained from the anti-crossing gap, the calculated  switching off point   coincide well the measurement results in Fig.(3).

\begin{figure}
\includegraphics[bb=0 0 612 530, width=4.25 cm, clip]{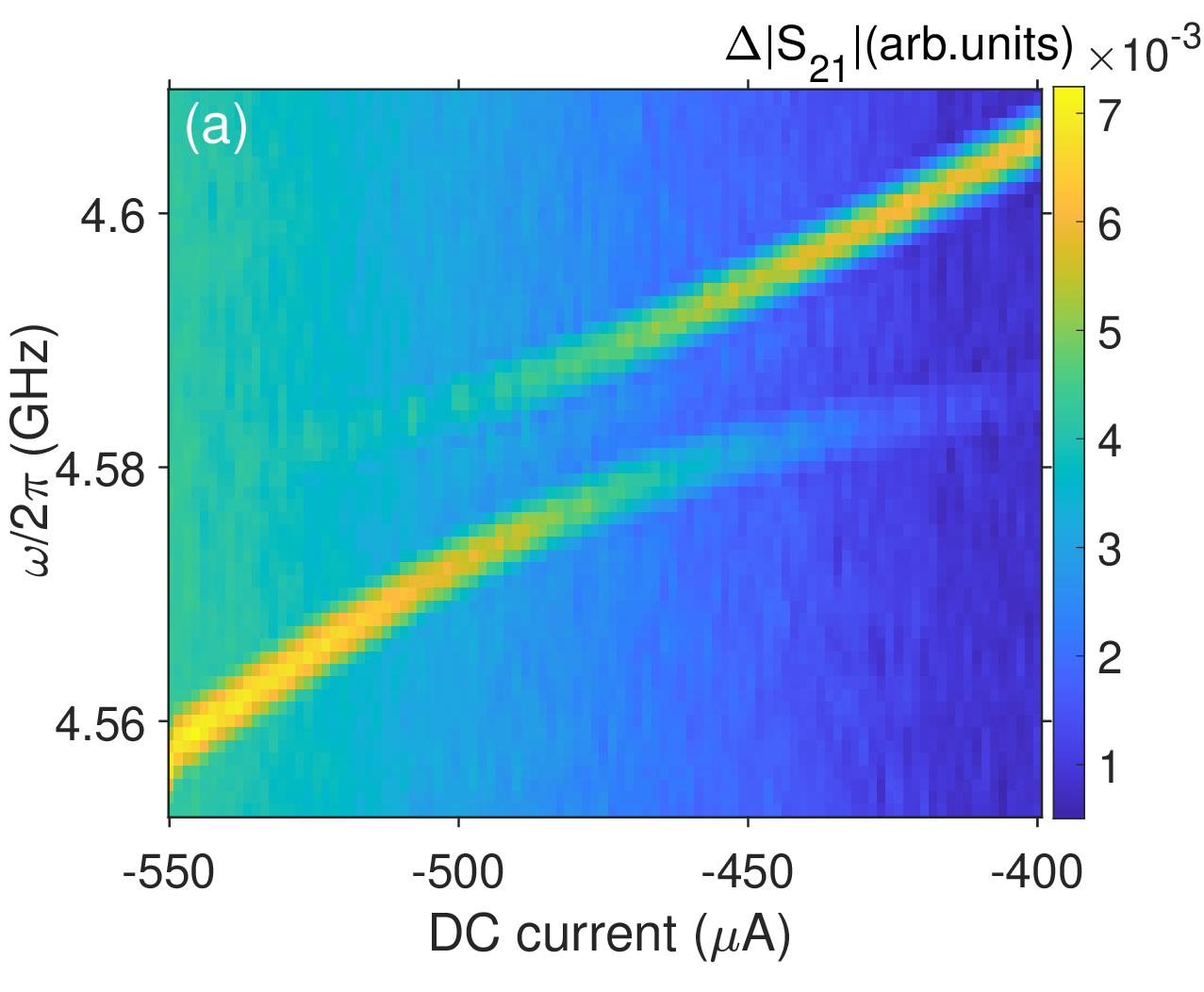}
\includegraphics[bb=0 0 610 530, width=4.25 cm, clip]{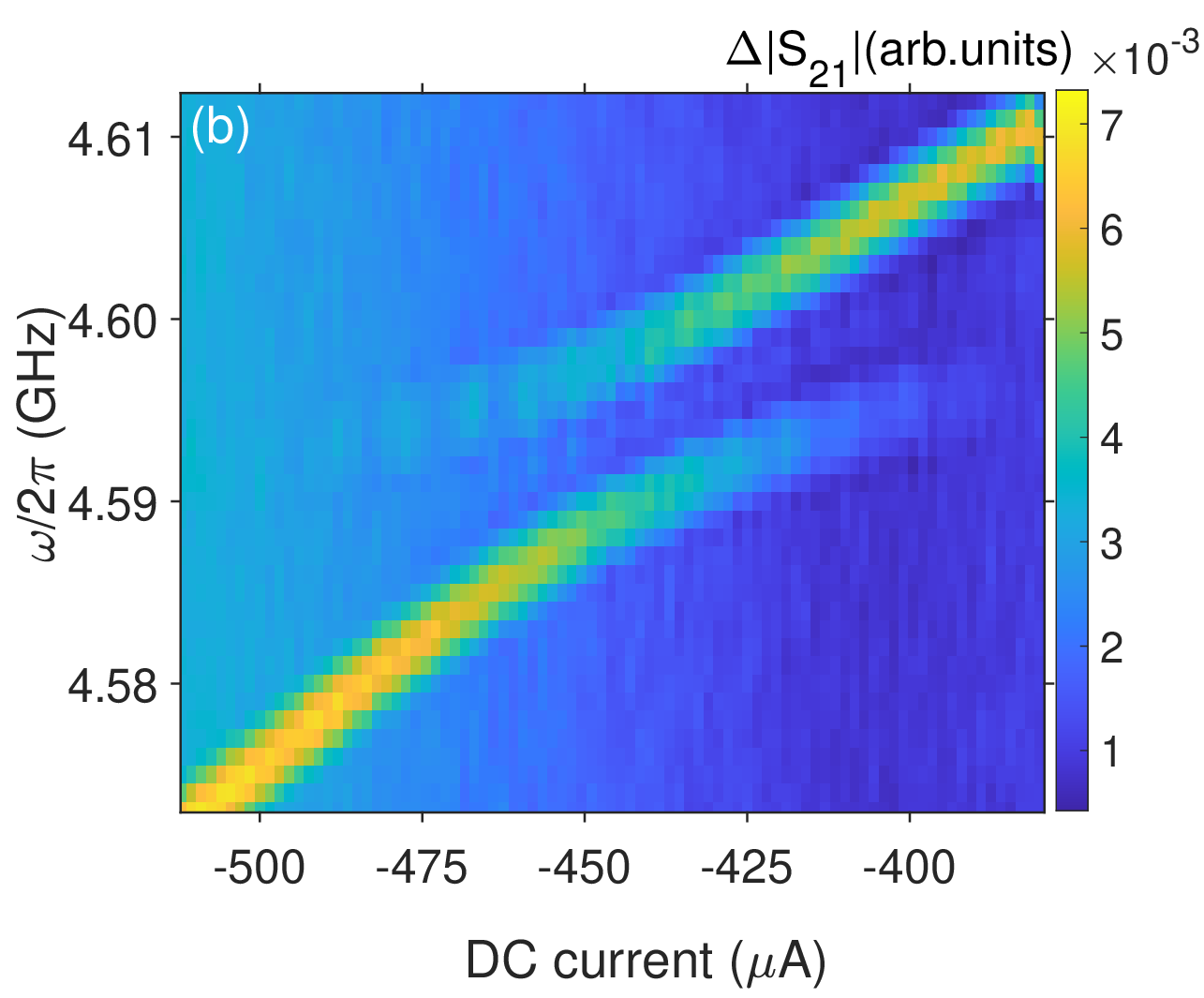}\\
\includegraphics[bb=0 10 630 510, width=4.20 cm, clip]{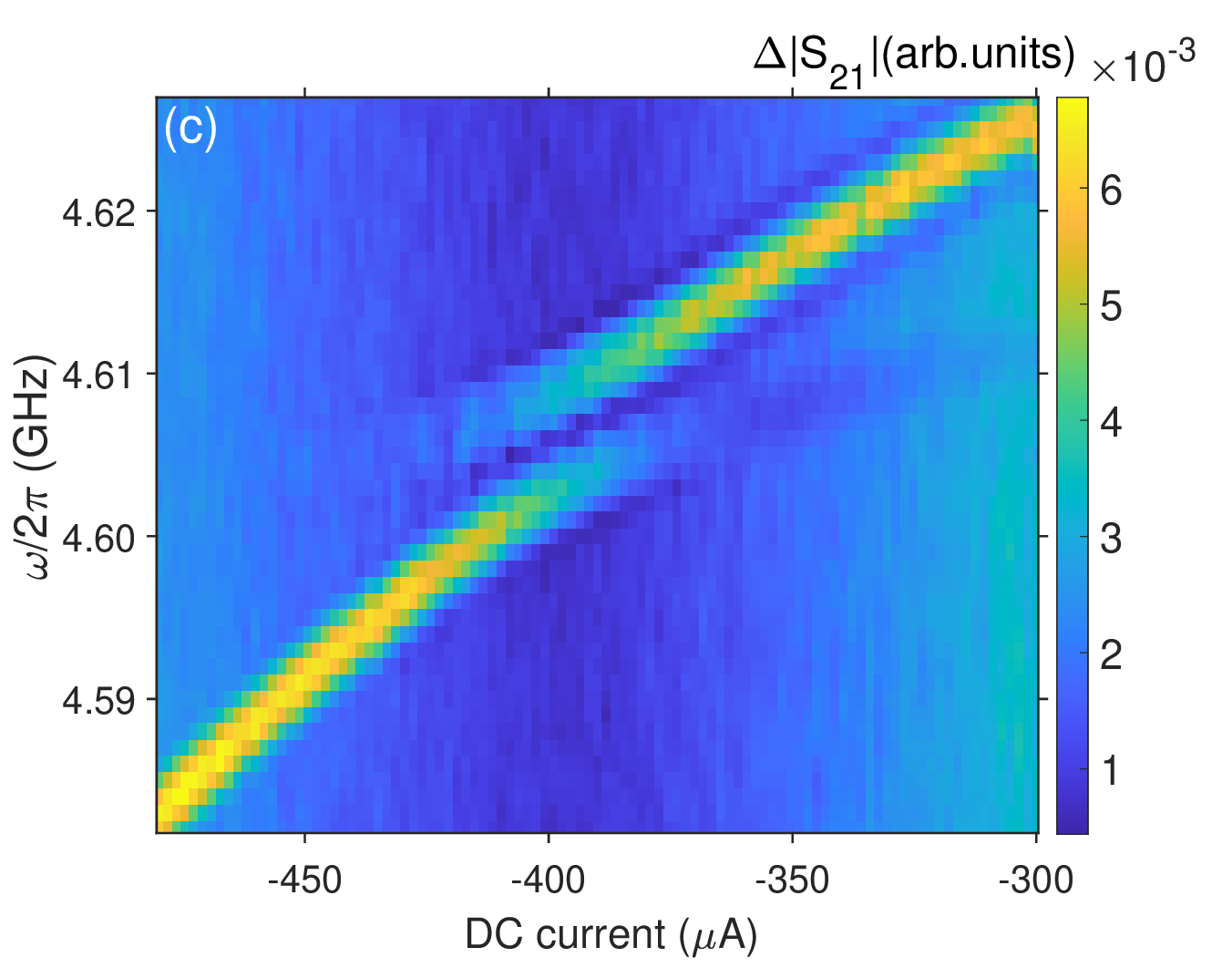}
\includegraphics[bb=10 10 640 510, width=4.20 cm, clip]{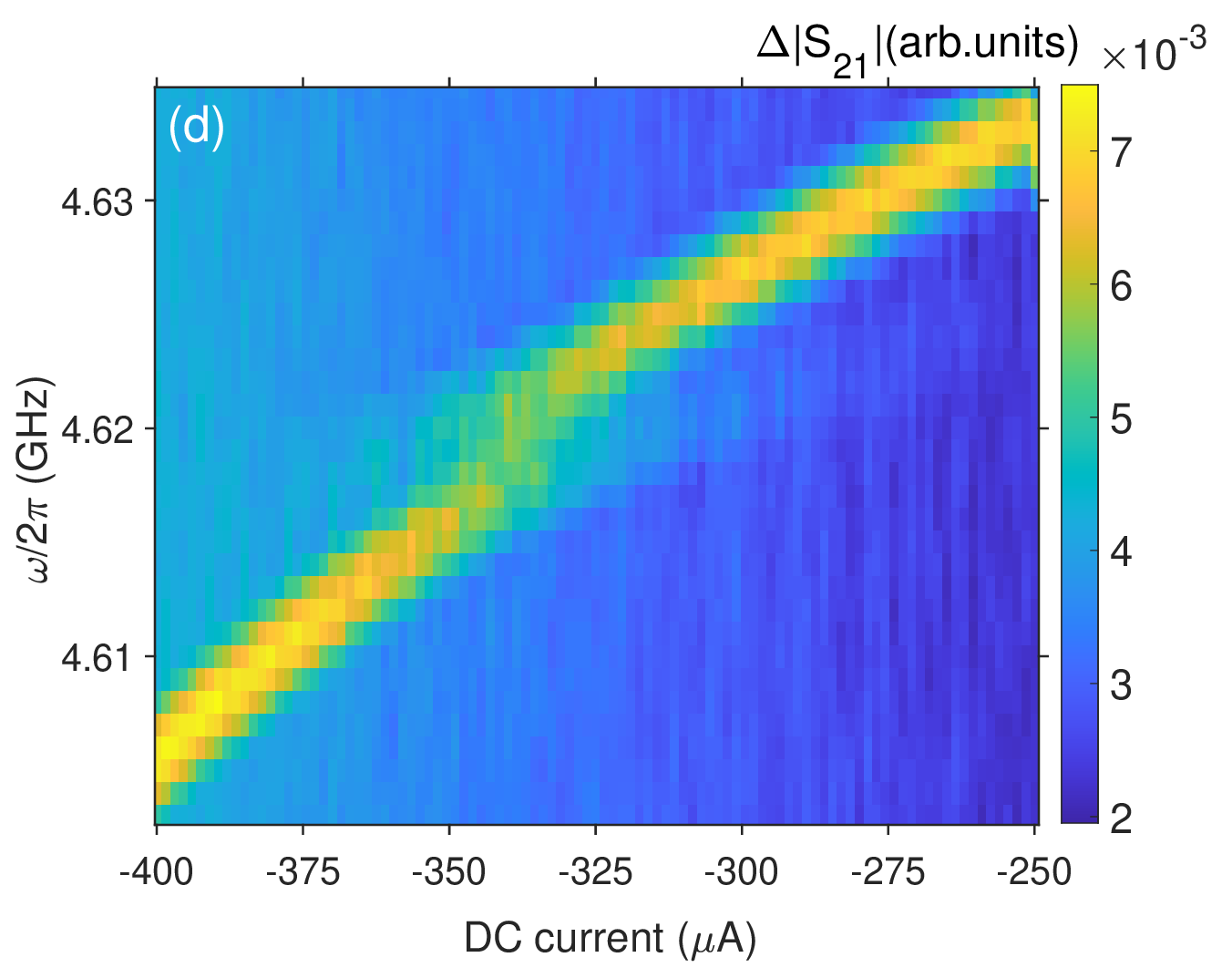}\\
\includegraphics[bb=10 10 610 500, width=4.25 cm, clip]{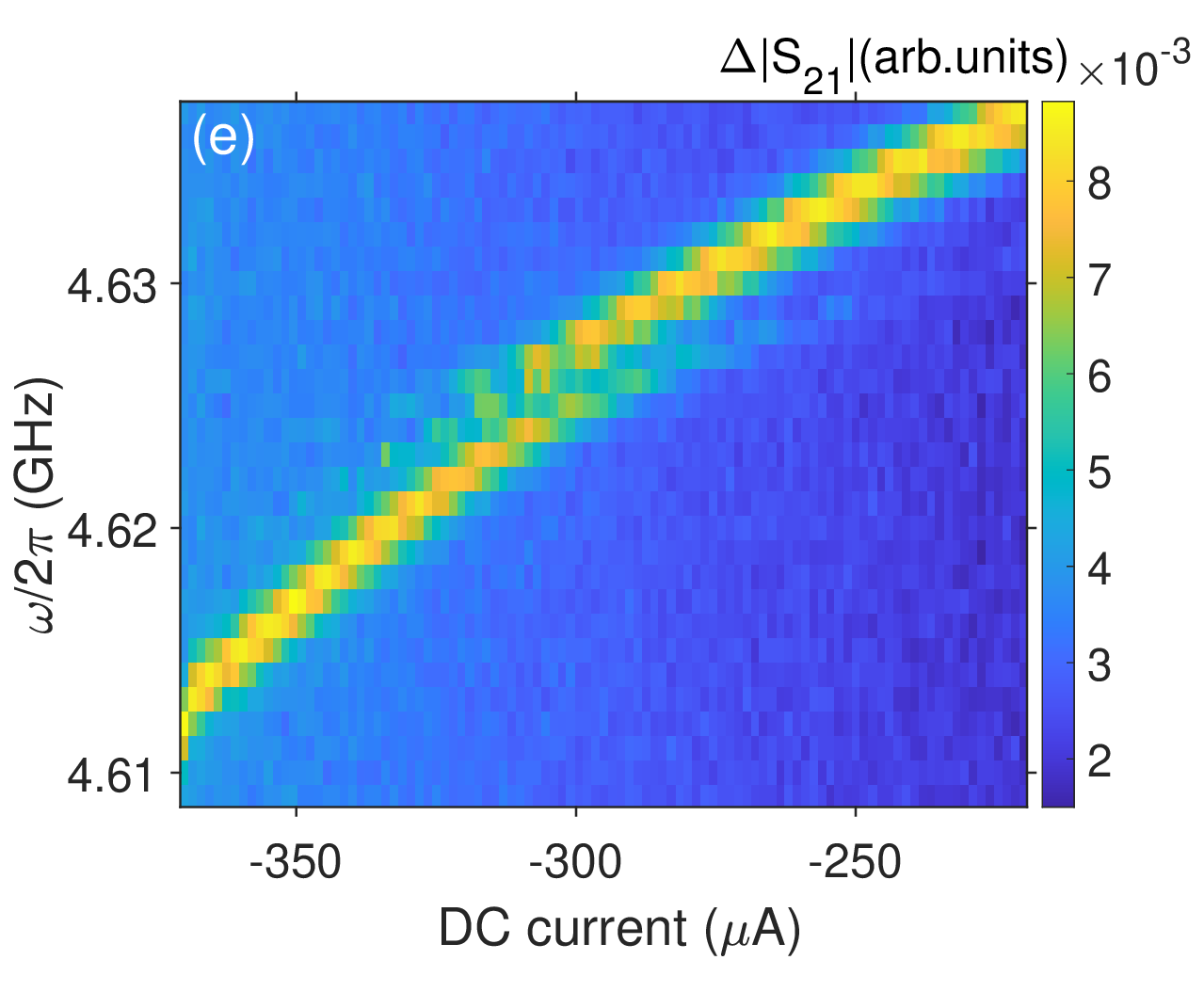}
\includegraphics[bb=10 10 610 500, width=4.25 cm, clip]{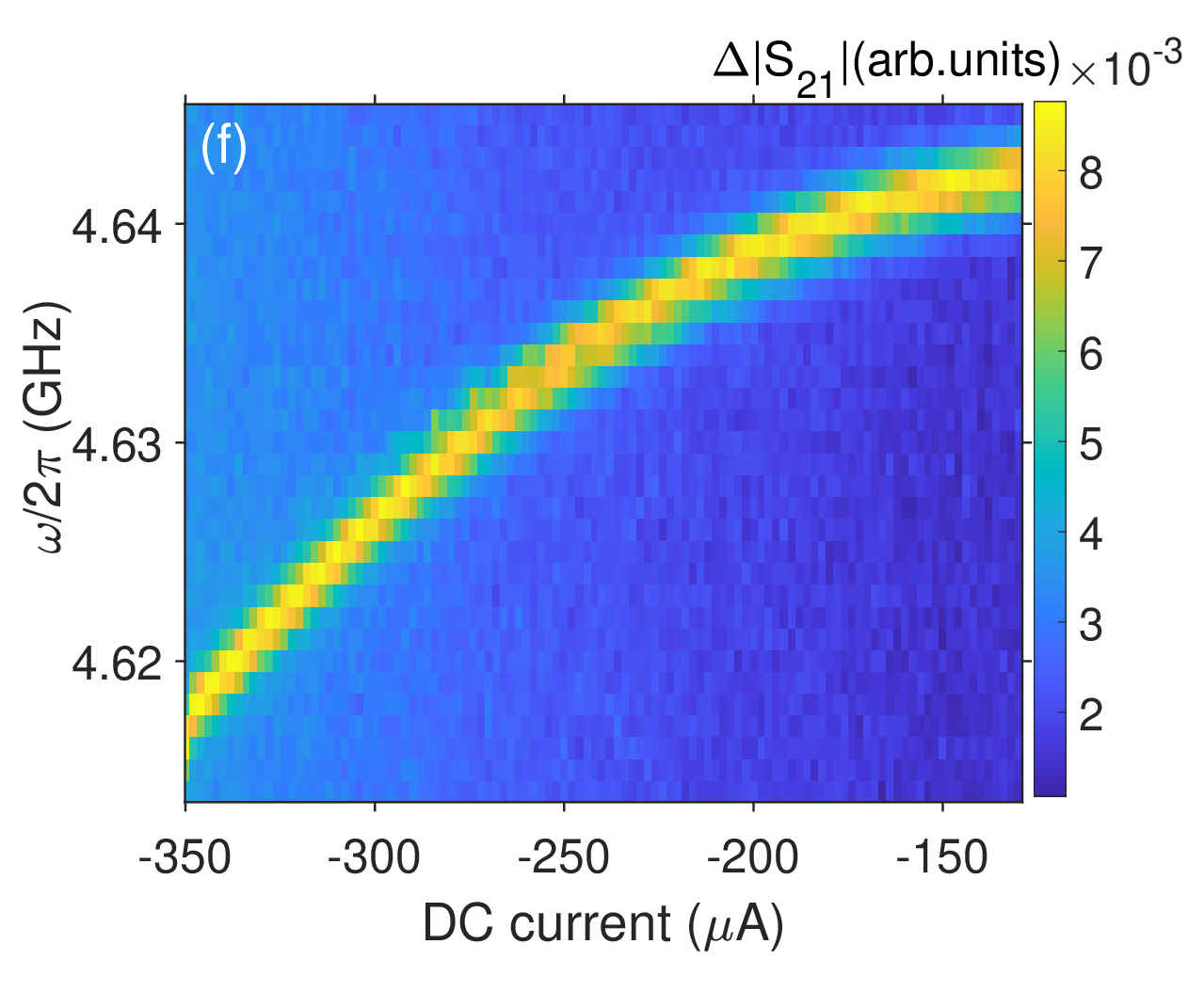}
\caption{(Color online) Tunable qubit-qubit anti-crossing gaps.
The qubit-2 is fixed at 6 different frequencies, and the anti-crossing gap are measured
by sweeping the frequency of qubit-1 (with the DC-biased current)
around  the corresponding  frequency of qubit-2 as shown in  (a)-(f).
}
\label{fig3}
\end{figure}

\begin{figure}
\includegraphics[bb=0 0 680 570, width=4.25 cm, clip]{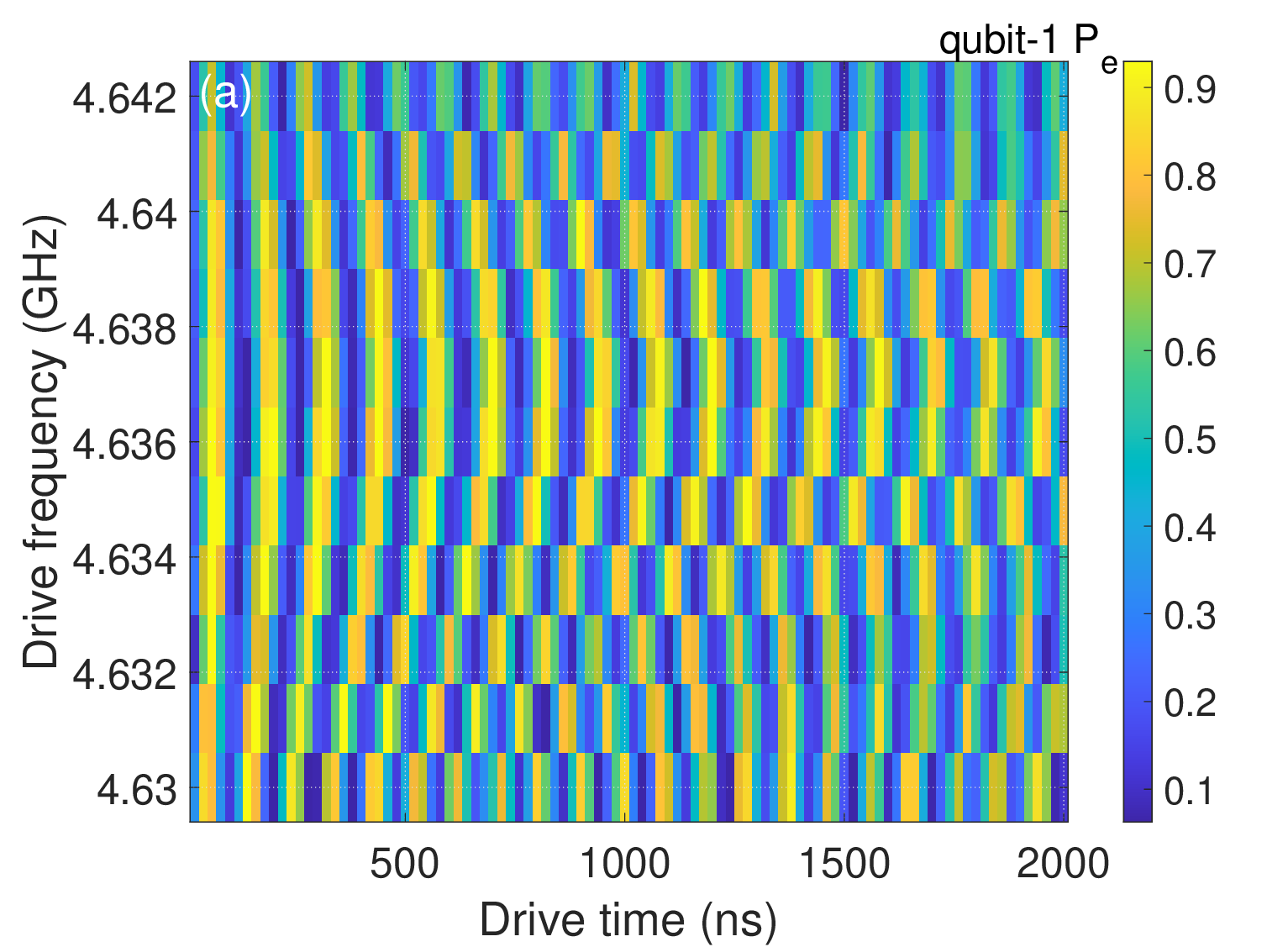}
\includegraphics[bb=0 0 680 570, width=4.25 cm, clip]{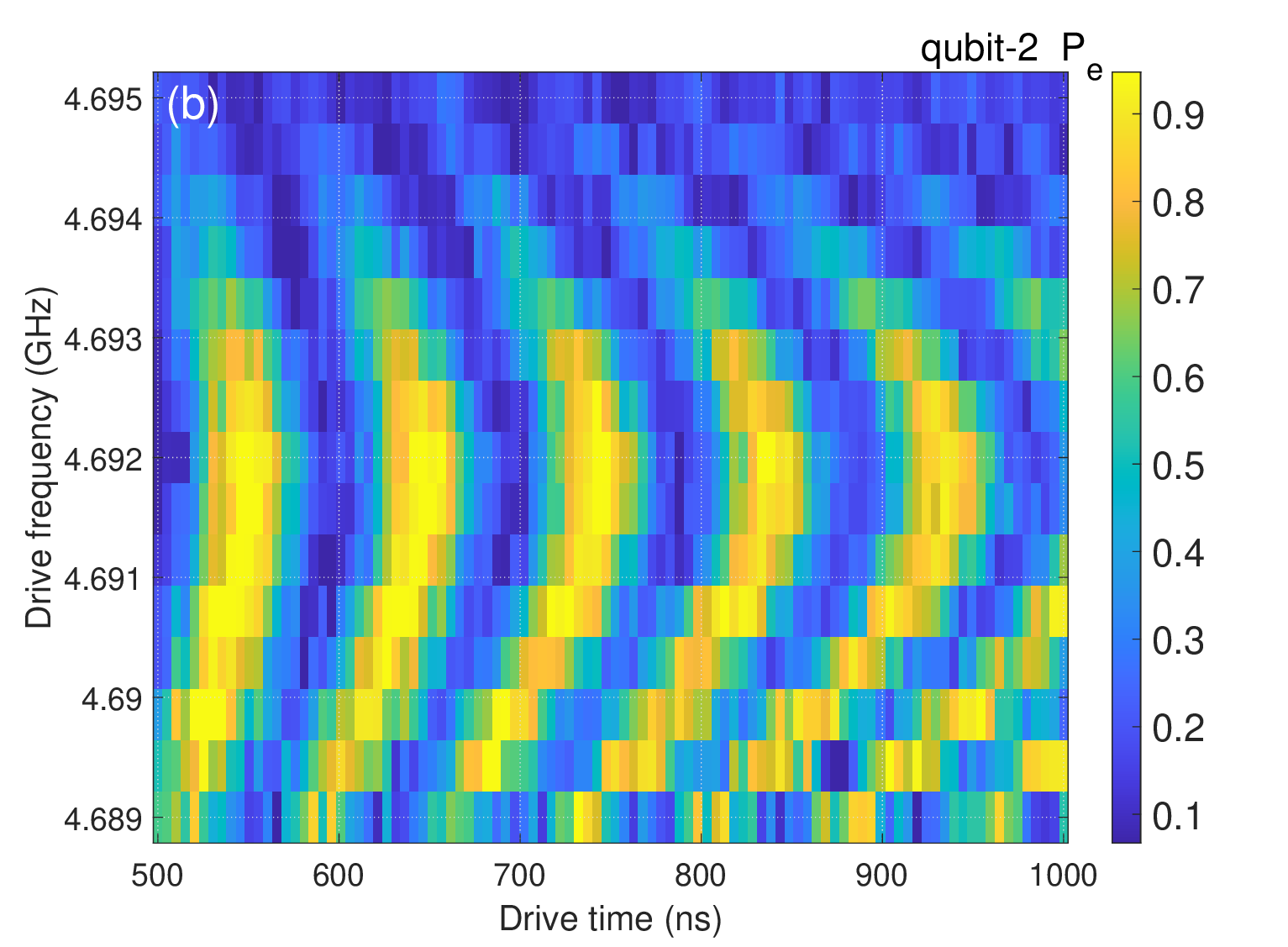}\\
\includegraphics[bb=0 10 550 500, width=4.2cm, clip]{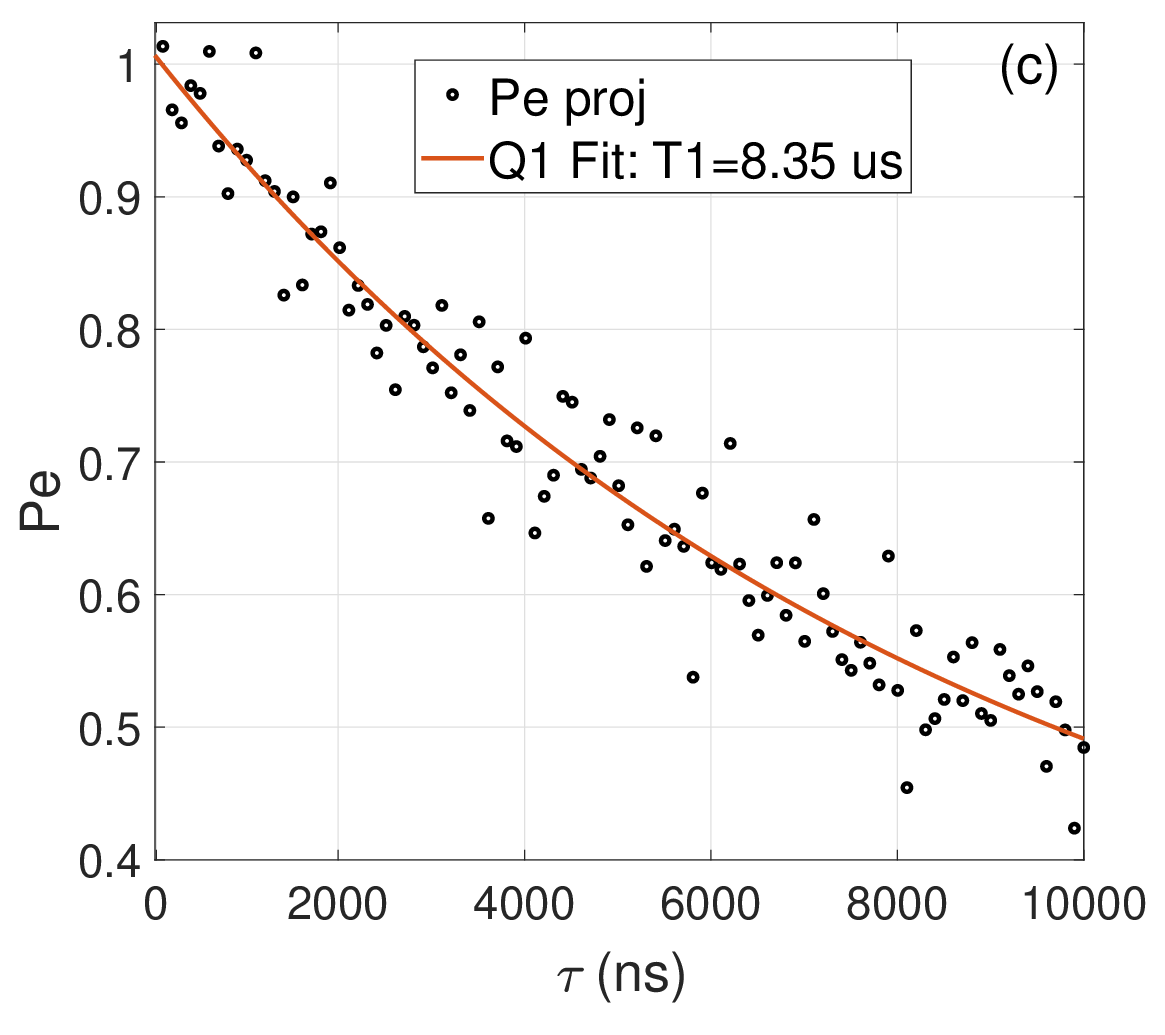}
\includegraphics[bb=0 5 550 500, width=4.2 cm, clip]{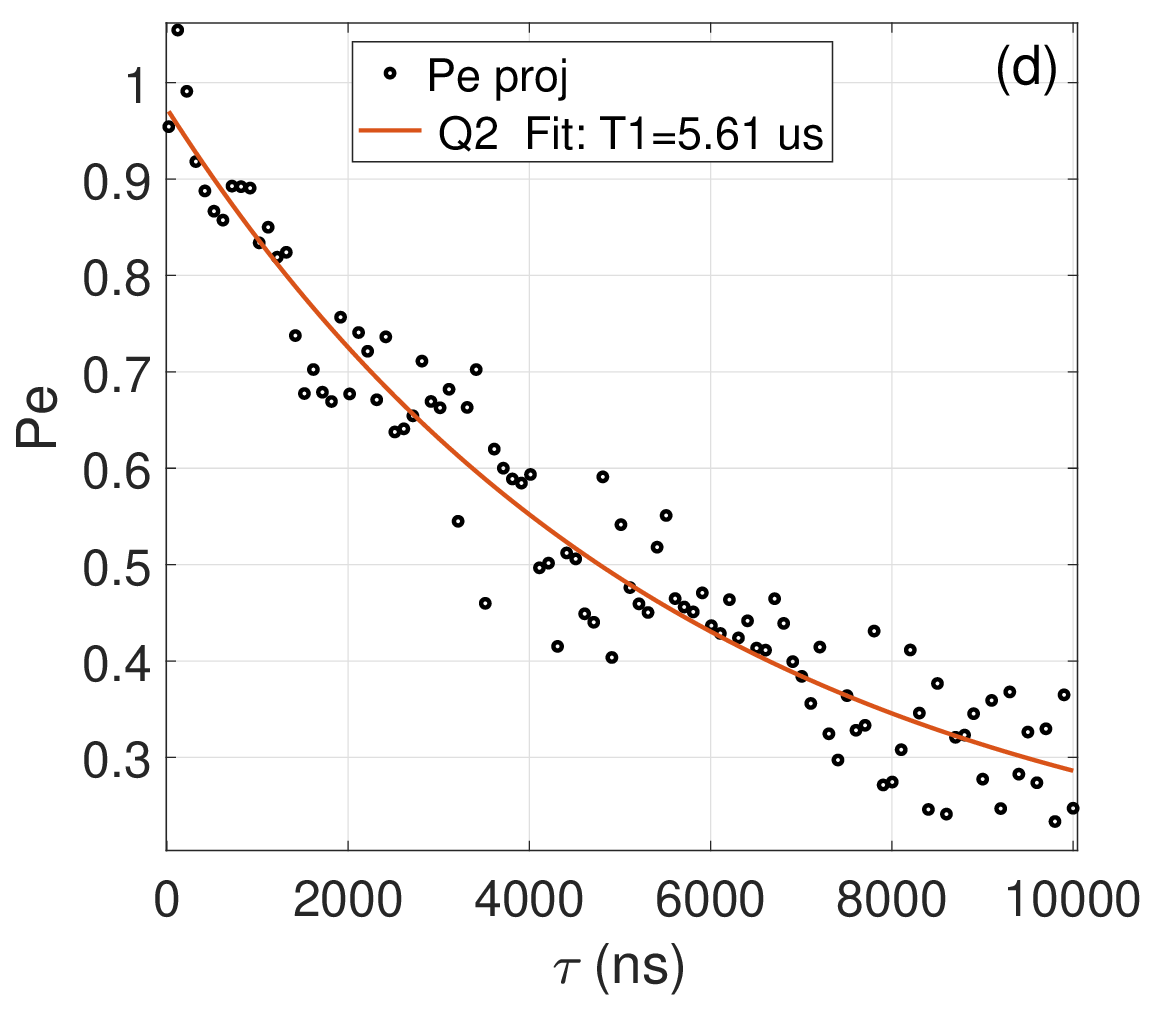}\\
\includegraphics[bb=0 0 520 470, width=4.2cm, clip]{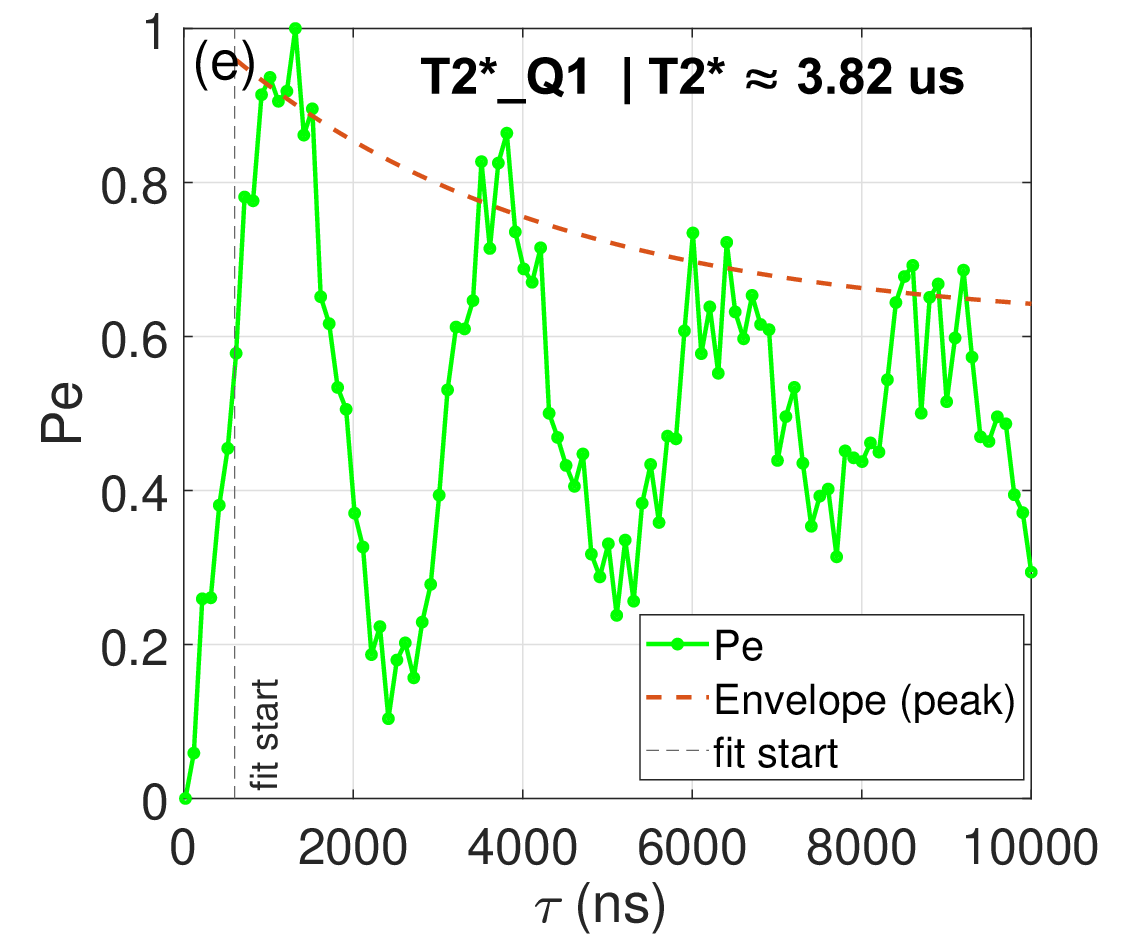}
\includegraphics[bb=0 0 520 470, width=4.2cm, clip]{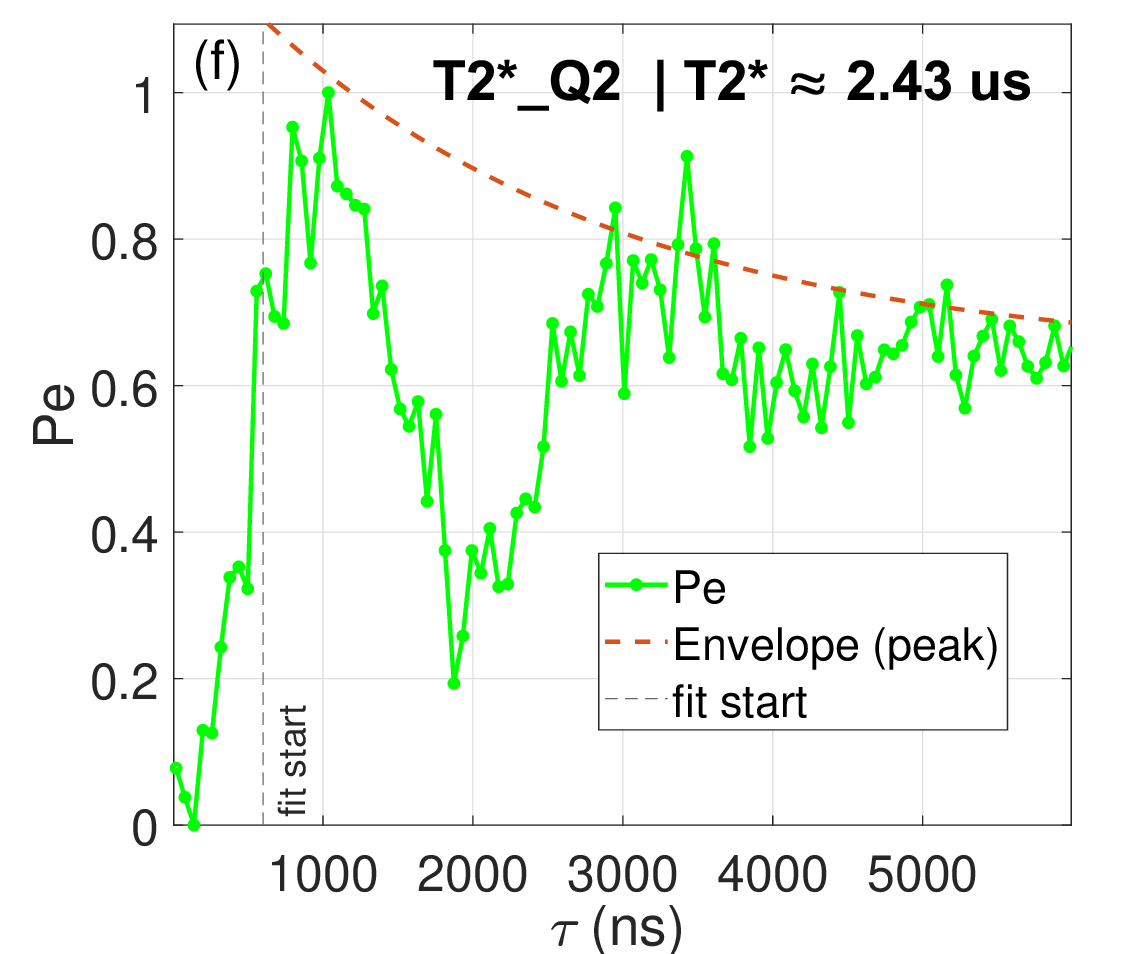}
\caption{(Color online) (a) Rabi oscillation of two qubits under the DC-bias current (no flux pulse).
The  center frequencies of two Rabi oscillation under are respectively $\omega^{(bias)}_1/2\pi=4.637$ GHz and $\omega^{(bias)}_2/2\pi=4.691$ GHz.
The $T_1$ (or $T_2$)  are respectably measured in (c) (or (e))  at 4.637 GHz  for qubit-\textbf{1} and  (d) (or (f)) at 4.692 GHz  for qubit-\textbf{2}.
}
\label{fig4}
\end{figure}

By only  shifting  50 MHz from the switching off points with flux pulse,  the anti-crossing gap
 effective qubit-qubit coupling can be tuned from 0 Hz to about 5 MHz as shown in Fig.3(a).
Thus the switching off positions  can be  very close to   two-qubit gate regimes (about 5 MHz effective qubit-qubit coupling),
thus the flux noises can be greatly suppressed during the quantum  operations for the double-resonator couplers superconducting quantum chip.
The double-resonators coupler can turn off the  qubit-qbuit coupling without  requiring
additional flux lines, this can  reduce  flux noises and the occupation number for
 high-frequency cables of dilution refrigerator.
The width of coupler resonator can be narrower in future design, than the size of resonator
   couplers can be much smaller space of the chip.
To show the value of qubit-resonator strength, the maximal frequency of one qubit is
 larger than the high frequency resonator-coupler which is not necessary in the
real superconducting quantum processor.

\section{Time-domain measurement}

The vacuum Rabi oscillation period can also reflect effective qubit-qubit coupling strength
between two qubits\cite{Liang,Wu2}. To avoid
 level crossing  between qubit-\textbf{2} and high-frequency resonator in the vacuum Rabi experiment,
  we tune qubit-\textbf{2} (or qubit-\textbf{1}) to 4.691 GHz (or 4.637 GHz) with the
  DC-bias currents.  DC-Biased Frequencies of two qubits are close to the switching off points (around 4.637 GHz as indicated by  Fig.2),
  this can reduce signal distortions  (at large  pulse amplitude) in the vacuum Rabi measurement.
   In the case of zero flux pulse, the  Rabi oscillation of two qubits are measured as shown in Figs.4(a) and 4(b).
In the absent of flux pulse, the coherent time $T_{1}$  and $T_{2}$ for two qubits
 are measured in Figs.4(c)-4(f) where the frequency detuning is about 50 MHz
 during the $T_1$ and $T_2$ measurement. The large fluctuations on the curves of  $T_1$ and $T_2$
   indicate the low readout  Signal-to-Noise Ratio (without Josephson parameter amplifier)
   and relative high   base temperature(above 25 mK) of dilution refrigerator.
 In the regimes of strong flux pulses, the Rabi response spectrum (phase) curve of qubit will be divergent
 as shown in Fig.8 (see Appendix).

 Pulse sequences employed for vacuum Rabi measurement are shown in Fig.5.
 In stage-\textbf{A}, the $\pi$ pulse drive qubit-2 to its first-excited state
 in the case of zero flux pulse (4.691 GHz) which reduce the frequency drift induced by the
    flux pulse distortion\cite{Yan,li1,Tian}.
Immediately after  $\pi$ pulse exciting qubit-\textbf{2} to its first excited states,
the flux pulse-\textbf{2} will tune  qubit-2  to 4.637 GHz in   stage-\textbf{B}, and simultaneously
    flux pulse-\textbf{1} shift qubit-\textbf{1} close to 4.637 GHz.
    The two near resonant qubits will exchange energy with
    each other in state-\textbf{B}, then  one dimensional vacuum Rabi curve
 can be plotted as the function  of   interaction duration $\tau$  as shown in Fig.6.
 The amplitude of flux pulse-\textbf{1} is  tunable to  sweep the transition frequency of qubit-1, and then we can
 measure a series of Vacuum Rabi curves at different qubit-qubit frequency detuning.
After the interaction finished(roughly stage-B), the qubits will
 return to the DC-biased frequency after zero the amplitudes of two flux pulse,
then we   read the occupation probabilities of qubit-\textbf{1}.
To reduce the signal drift, the time interval between driving $\pi$-pulse and readout pulse
fix  at a certain value (about 1.2us) smaller than the coherent times of two qubits.

\begin{figure}
\includegraphics[bb=0 0 500 290, width=7.5 cm, clip]{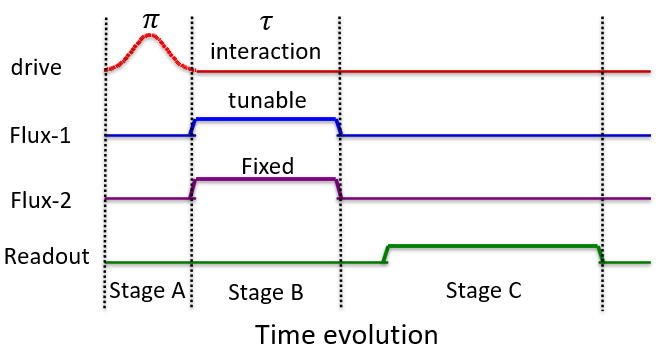}
\caption{(Color online) Pulse sequence employed for the Vacuum Rabi measurement.
 In the stage A, the $\pi$ pulse drive qubit-\textbf{2} to its first-excited state.
In the stage B,  the flux pulse-\textbf{2} tune the qubit-\textbf{2}  to 4.637 GHz,
simultaneously qubit-\textbf{1} is tuned close to 4.637 GHz with the  flux pulse-\textbf{1}.
The two qubits interaction interval ($\tau$) is decided
by simultaneous holding period of two flux pulses in  stage-B.
After the interaction finished, two flux pulse turned off, and the quantum states of qubit-\textbf{1}
will be readout in stage C (at frequency 4.637 GHz).
The time interval between driving $\pi$-pulse and readout pulse
fixes  at a certain value (within the coherent times of two qubits.)
}.
\label{fig5}
\end{figure}

For the vacuum Rabi oscillation measurement, the qubit-\textbf{1}
exchange energy with qubit-\textbf{2} through near resonant interactions, then the occupation
probabilities of qubit-\textbf{1} with interaction time duration\cite{Liang,Wu2}
In the case of two qubits near resonant, the  maximal occupation probability  should
 be $P_{1,max}\sim \exp{(-T_{swap}/T_2)}$, while $T_{swap}=\pi/2|g_{eff}|$. As indicated by the small
  anti-crossing gaps in Fig.2, the effective  qubit-qubit coupling is quite weak
 close to the switching off point.
If two qubits are in the  off-resonant case, the maximal occupation probabilities of qubit-\textbf{1}
first excited-state is  $P_{1, max}=4g^2_{eff}/(\Delta^2_{12}+4 g^2_{eff})$ which is usually smaller than one,
 with $\Delta_{12}=\omega_{2}-\omega_{1}$.
 Since $T_{swap}$ is close and even larger than  $T_2$ in this measurement,
thus the maximal occupation probabilities  for qubit-\textbf{1} can be much smaller than one as indicated by the simulation
 result with the lindblad master equation (see Fig.6(a))\cite{Johansson1,Johansson2}.

\begin{figure}
\includegraphics[bb=20 0 550 470, width=4 cm, clip]{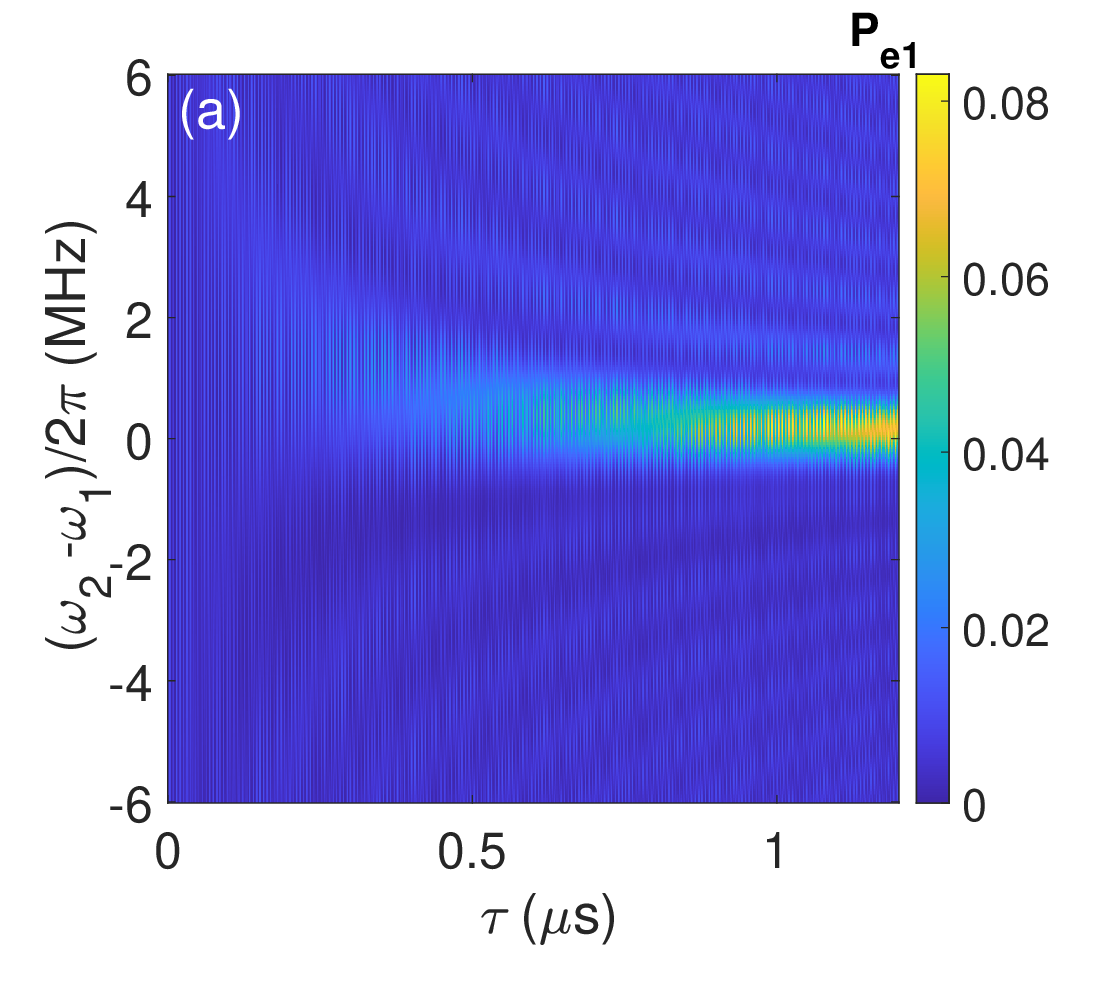}
\includegraphics[bb=0 0 450 480, width=3 cm, clip]{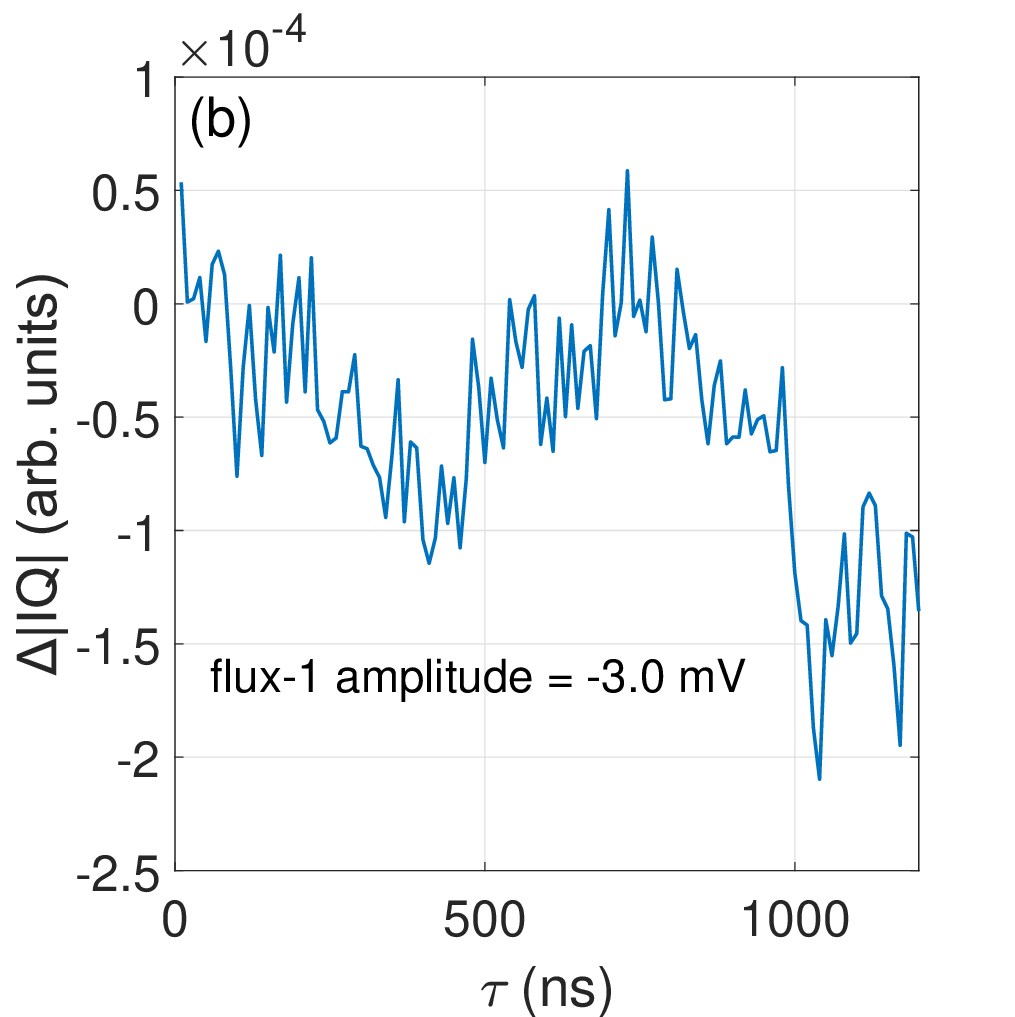}\\
\includegraphics[bb=0 0 450 480, width=2.75 cm, clip]{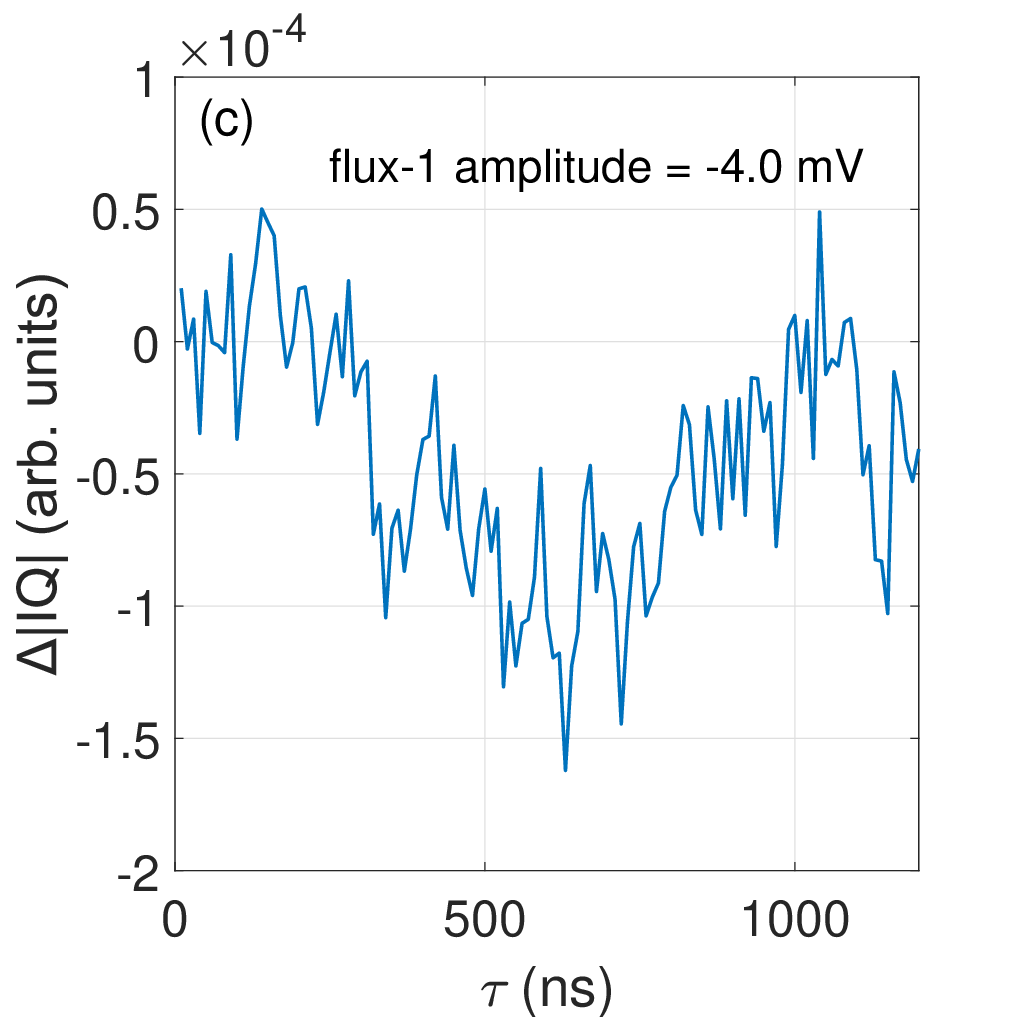}
\includegraphics[bb=0 0 450 480, width=2.75 cm, clip]{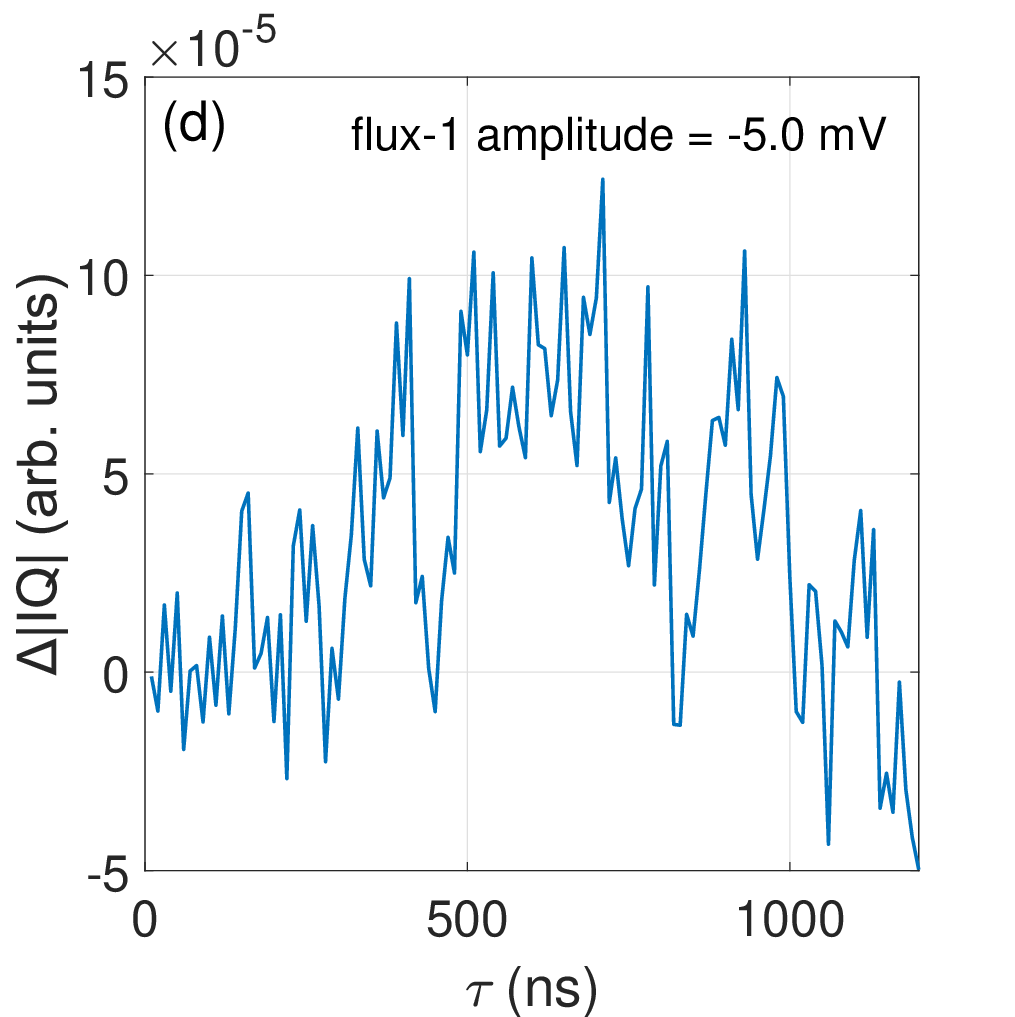}
\includegraphics[bb=0 0 450 480, width=2.75 cm, clip]{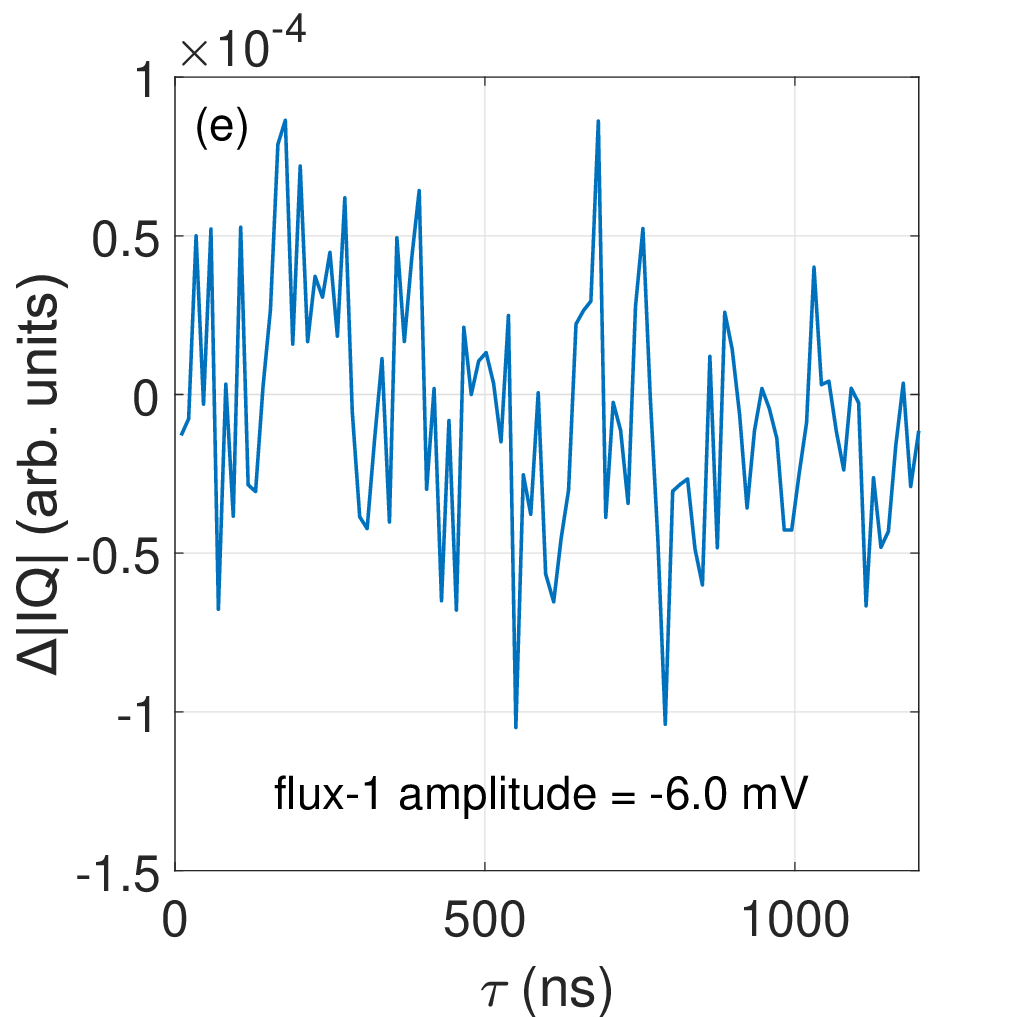}\\
\includegraphics[bb=0 0 450 480, width=2.75 cm, clip]{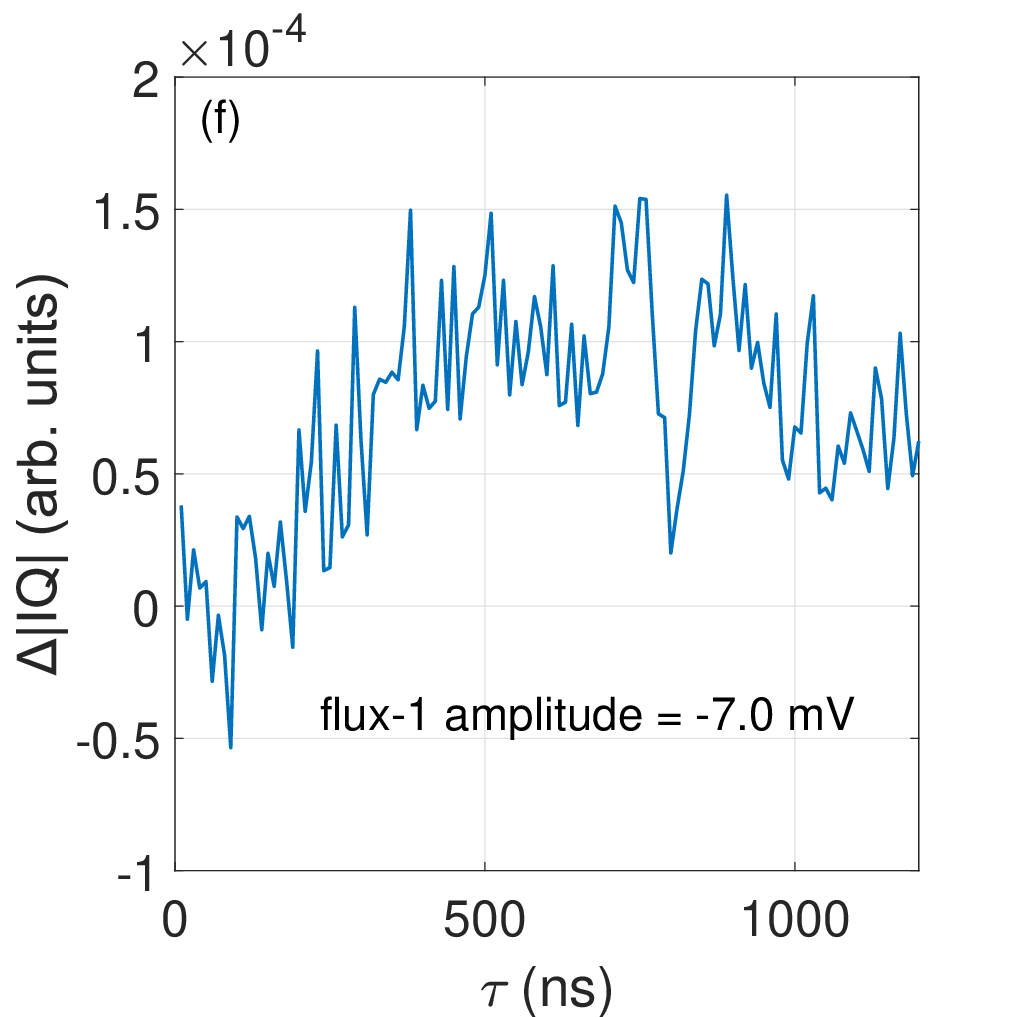}
\includegraphics[bb=0 0 460 480, width=2.75 cm, clip]{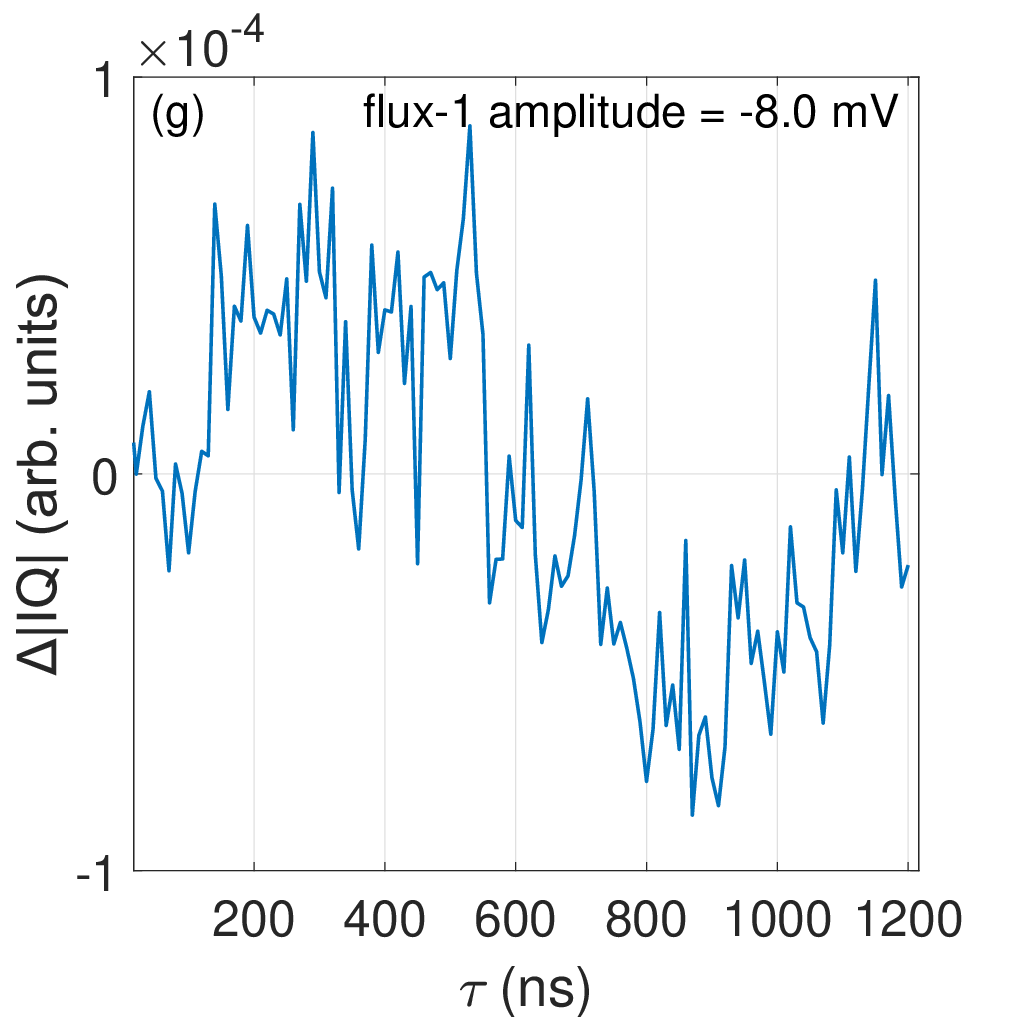}
\includegraphics[bb=0 0 450 480, width=2.75 cm, clip]{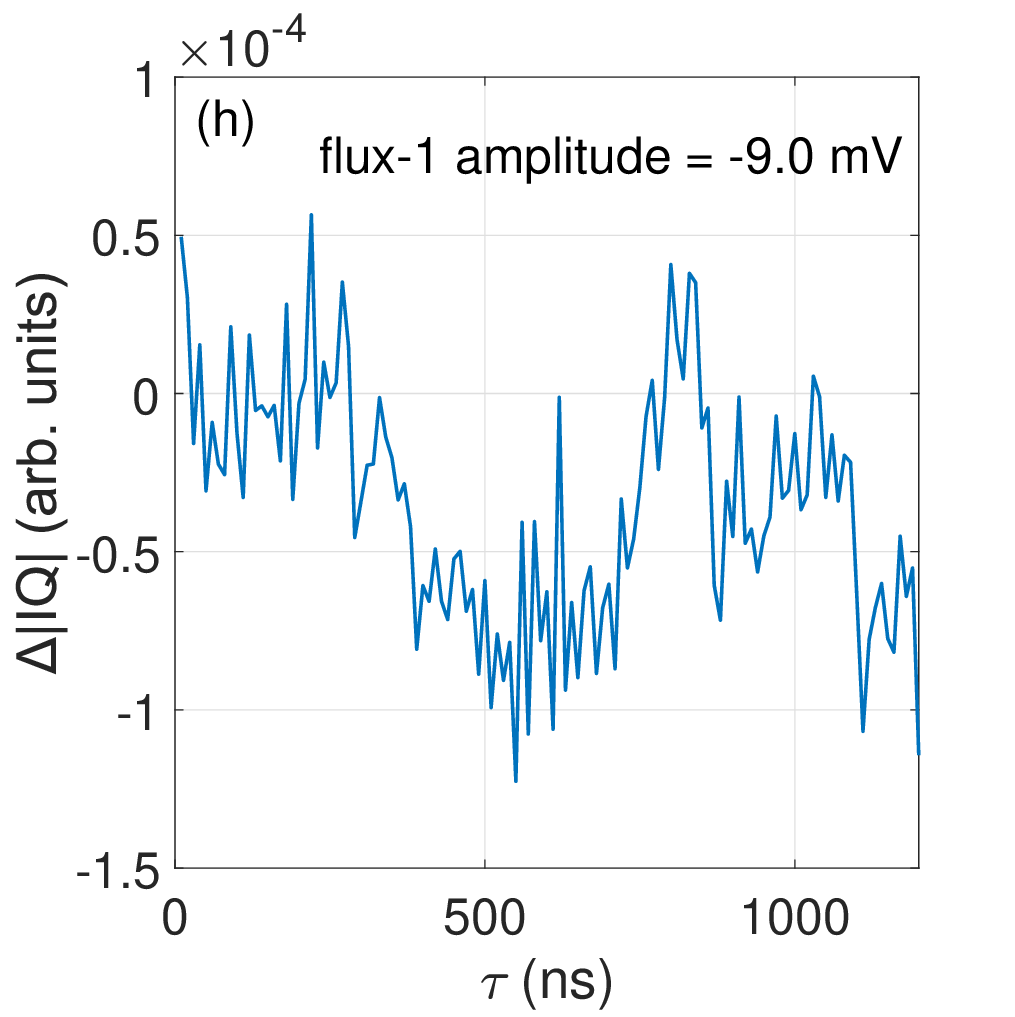}
\caption{(Color online) Vacuum Rabi oscillation. (a) Simulation of Vacuum Rabi oscillation.
 The occupation probabilities of qubit-1 as the function of
 interaction time and frequency detuning between two qubits. (b)-(h)  Readout contrast  relative to
 the ground-state baseline ($|IQ|-baseline$) at different frequency detuning between two qubits.
The   pulse amplitude for qubit-1 vary  in (b)-(h),
and the flux-1 amplitude=:(b)-3 mV;(c)-4 mV;(d)-5 mV;(e)-6 mV;
(f)-7 mV; (g) -8 mV; (g)-9 mV.
}
\label{fig6}
\end{figure}

 In the absence of Josephson parameter amplifier and the relative high base temperature (above 25mK),
 the Signal-to-Noise Ratio is not high in this measurement. If we use the normalization method to calculate
      excited state occupation probabilities,  more errors will be induced by the noise data.
 As shown in Figs.6(b)-6(h),  we use  $\Delta|IQ|(=|IQ|-baseline)$ to describe  the  variation of  the readout signal
which can   reflect envelope variations of vacuum Rabi oscillation.
Even with low SNR, we still can see some feature of envelope variation  of $\Delta|IQ|$
from the noisy data (see Figs.6(b)-6(h)).
When the flux pulse amplitude is close to -6.0  mV, the envelope of $\Delta|IQ|$ is not clear in Fig.6(e), which should be
 close to switching off point(see fig.3). As the change of the  flux pulse   amplitude of qubit-\textbf{1},
 the envelope for the curve of $\Delta|IQ|$ become clearer at larger frequency detuning ($|\Delta_{12}|$)
  because of nonzero energy exchange between two  qubits, the oscillation periods of  $\Delta|IQ|$
   reflect variation of  qubit-qubit coupling strength.
 The switching off point  measured by the Vacuum Rabi oscillation roughly coincide with the anti-crossing gap in the frequency domain.
  The measurement result should be enhanced if  the Josephosn parameter amplifier, Purcell filter, Z-pulse distortions are applied in future experiment.

 \section{Conclusions}\label{conclusion}

In conclusion, we  experimentally studied tunable qubit-qubit  coupling in the double-resonator coupler type
 superconducting quantum circuit. The induced indirect qubit-qubit coupling by two resonators can be cancelled,
 so the switching off can be realized without the direct qubit-qubit coupling.
In the frequency domain and time-domain, we observed the variation of the effective qubit-qubit coupling tuning by frequency detuning between the qubits and resonator couplers.
By reducing the designed line widths of resonator-coupler, than occupation space of  resonator couplers can be much smaller in the multi-qubit superconducting quantum chips.

\section{ACKNOWLEDGMENTS}

Hui Wang thank Zhiguang Yan for
valuable suggestions on the  hardware setup,  codes writing, and   result analysis of the low-temperature measurments.
We also thank Shiyu Wang and  Russell S. Deacon for friendly help during the measurements and fabrications of the superconducting samples.
 J.S. Tsai is supported by the Japan Science
and Technology Agency (Moonshot R D, JPMJMS2067;
CREST, JPMJCR1676) and the New Energy and Industrial Technology
 Development Organization (NEDO, JPNP16007).

\section*{Appendix A: The numerical calculation of the energy levels}

\setcounter{equation}{0}
\renewcommand\theequation{A.\arabic{equation}}

 \subsection{Experiment setup}

\begin{figure}
\includegraphics[bb=0 0 800 520, width=9 cm, clip]{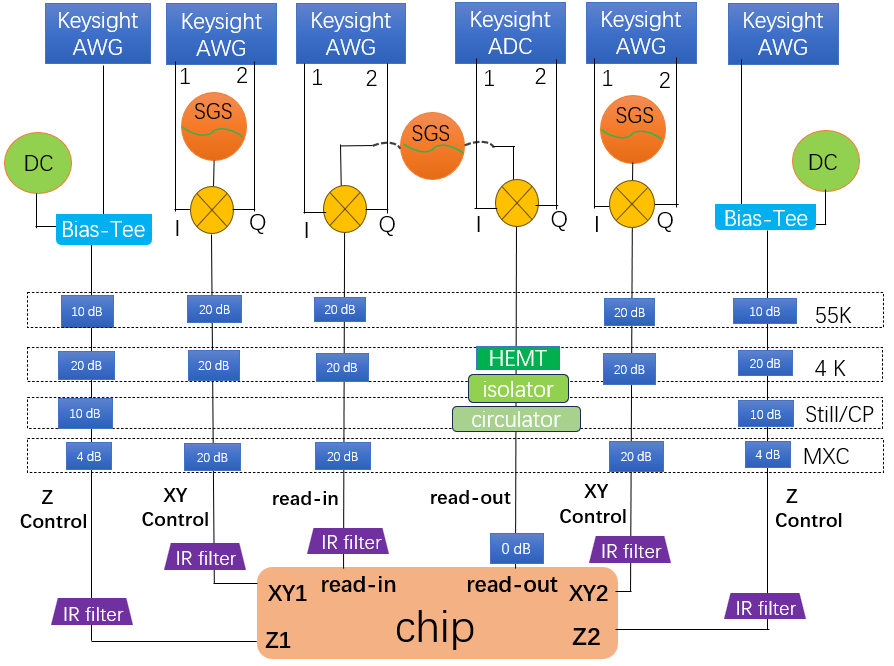}
\caption{(Color online)  Electronics and sample schematic diagram for the vacuum rabi measurement.
}
\label{fig7}
\end{figure}

The measurement is use the keysight M3202A and digitizer M3102A, the microwave signals are created by the RS sgs100a.
The attenuation in the read-in and XY control line are about 60db, and about 44db for the Z-control line
The DC-current and the flux pulse is combined by the bias-Tee in room temperature. The  infrared filters are installed for the driving and readout cables.
The DC current will bias the qubits at suitable frequency points for the Rabi or
the vacuum Rabi measurement, while the flux pulse is for the fast sweeping
 of qubits' frequency, and they  combined together with each other in room temperature with a bias-Tee.

\subsection{Rabi-frequency spectrum under flux-pulse}

Close to the DC-bias frequency of  qubit-\textbf{1} (qubit-\textbf{2}) at 4.691 GHz (or 4.637 GHz), the Rabi response
 spectrum of two  qubits can be measured by applying corresponding flux pulses are shown in Fig.8.
 The energy spectrum curves of two qubit become divergent at large   amplitude of flux pulse
  because of Z pulse distortion\cite{Yan,li1,Tian}.   We
 did not calibrate the Z-pulse distortion in this experiment, and the vacuum Rabi measurement will focus
 on the parameter  regimes close to 4.637 GHz where  the
   Z-pulse distortion is small because of  small flux pulse amplitudes.

\begin{figure}
\includegraphics[bb=0 0 500 420, width=4.25 cm, clip]{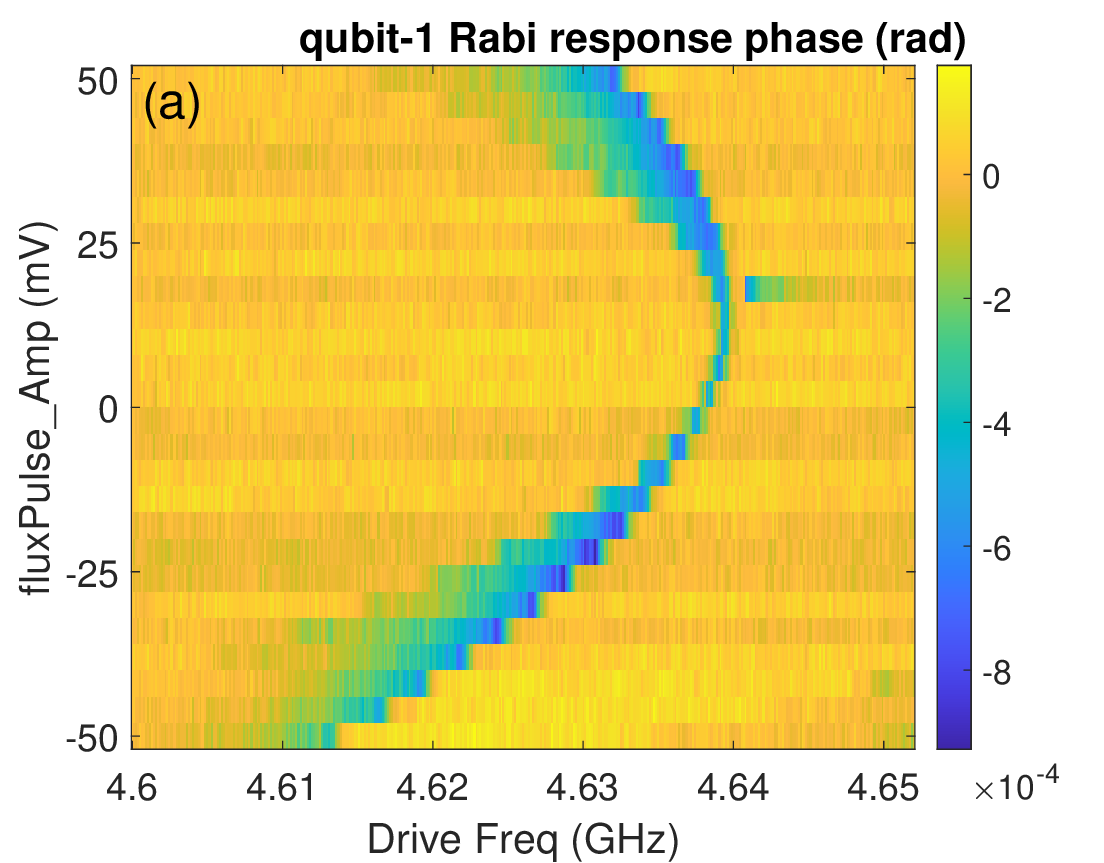}
\includegraphics[bb=0 0 500 420, width=4.25 cm, clip]{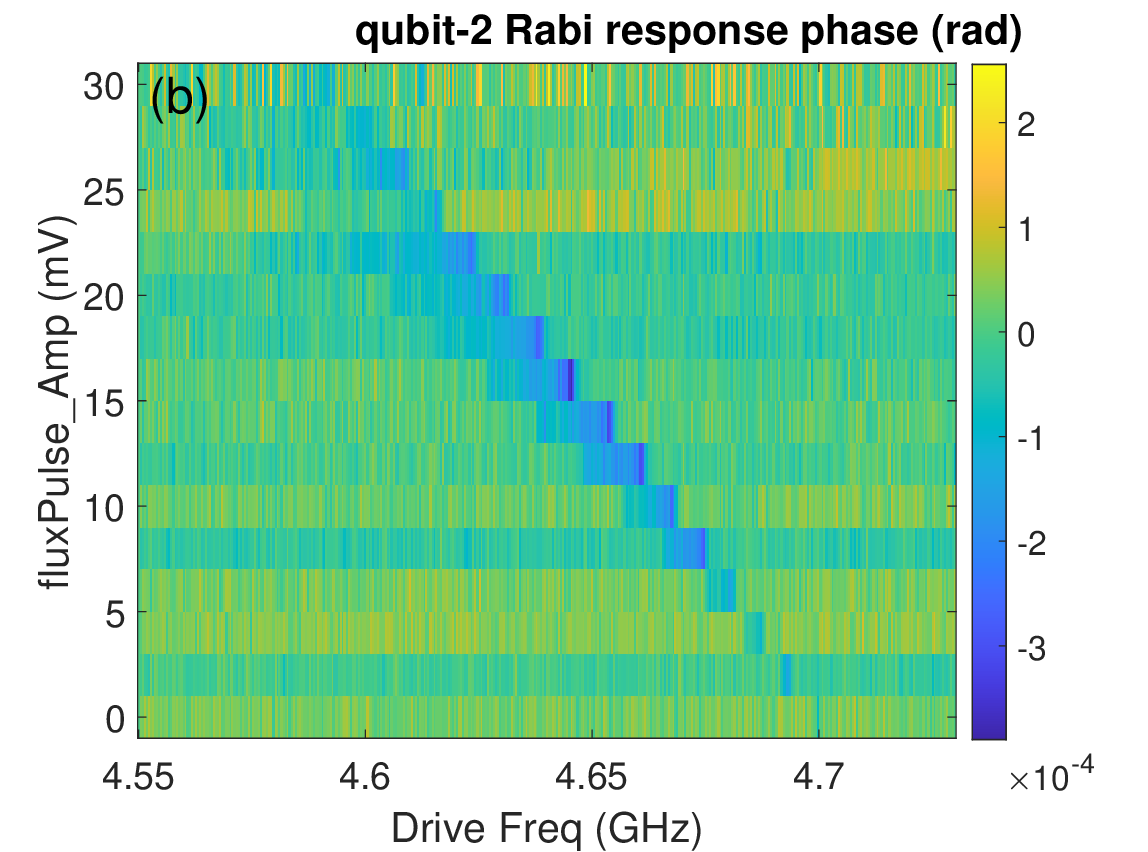}
\caption{(Color online) The Rabi response spectrum (phase).
Rabi response spectrums (phase) for (a) qubit-\textbf{1} and
(b) qubit-\textbf{2} under the flux pulses.
The zero flux correspond to the DC-bias frequencies of two qubits.
Rabi response spectrums become divergent
at large pulse amplitude because of  Z-pulse distortion.
}
\label{fig8}
\end{figure}

\end{document}